\documentclass[]{aa}
\usepackage{graphicx}
\usepackage{rotating,subfigure,amssymb,afterpage}
\usepackage{txfonts}
\usepackage{natbib}
\usepackage[pdftitle={The {\sf REFLEX II} Galaxy Cluster Survey},
pdfauthor={H. B\"ohringer et al.},
pdfsubject={}, pdfkeywords={}, bookmarks, bookmarksopen, 
colorlinks=true, linkcolor=blue, citecolor=blue, anchorcolor=blue,
urlcolor=blue]{hyperref} 
\newcommand{\propsim}{\lower 3pt \hbox{$\, \buildrel {\textstyle
      \propto}\over {\textstyle \sim}\,$}}
\begin{document}
   \title{The extended ROSAT-ESO Flux Limited X-ray Galaxy Cluster
   Survey (REFLEX II)\\ II. Construction and Properties of the Survey
\thanks{Based on observations at the European Southern Observatory La Silla,
   Chile}}

   \author{Hans B\"ohringer\inst{1}, Gayoung Chon\inst{1}, Chris A. Collins\inst{2}, 
         Luigi Guzzo\inst{3}, Nina Nowak\inst{4}, Sergei Bobrovskyi\inst{5}}

   \offprints{H. B\"ohringer, hxb@mpe.mpg.de}

   \institute{$^1$ Max-Planck-Institut f\"ur extraterrestrische Physik,
                   D-85748 Garching, Germany.\\
              $^2$ Astrophysics Research Institute, Liverpool John Moores University, 
                    Twelve Quays House, Egerton Wharf, Birkenhead,  CH41 1LD, UK\\
              $^3$ INAF - Osservatorio Astronomico di Brera-Merate, via Bianchi 46,
                   I-23807 Merate, Italy\\
              $^4$ Max-Planck-Institut f\"ur Physik, F\"ohringer Ring 6,
                   D-80805 M\"unchen, Germany\\
              $^5$ Deutsches Zentrum f\"ur Luft-und Raumfahrt, 
                     Oberpfaffenhofen, D-82234 Weßling, Germany}

   \date{Submitted 15/08/12}

\abstract
{Galaxy clusters provide unique laboratories to study
astrophysical processes on large scales and are important probes for cosmology. 
X-ray observations are currently the best means of detecting and characterizing
galaxy clusters. Therefore X-ray surveys for galaxy clusters are one of the
best ways to obtain a statistical census of the galaxy cluster population.}
{In this paper we describe the construction of the REFLEX II galaxy cluster
survey based on the southern part of the ROSAT All-Sky Survey. REFLEX II extends
the REFLEX I survey by a factor of about two down to a flux limit 
of $1.8 \times 10^{-12}$ erg s$^{-1}$ cm$^{-2}$ (0.1 - 2.4 keV).}
{We describe the determination of the X-ray parameters, the process of X-ray source
identification, and the construction of the survey selection function.}
{The REFLEX II cluster sample comprises currently 915 objects.
A standard selection function is derived for a lower source count limit
of 20 photons in addition to the flux limit. 
The median redshift of the sample is $z = 0.102$.
Internal consistency checks and the comparison to several other galaxy cluster 
surveys imply that REFLEX II is better than 90\% complete with a contamination 
less than 10\%.}
{With this publication we give a comprehensive statistical description of the
REFLEX II survey and provide all the complementary information necessary 
for a proper modelling of the survey for astrophysical and cosmological 
applications.}

\authorrunning{B\"ohringer et al.}
\titlerunning{The {\sf REFLEX II} Galaxy Cluster Survey}
   \maketitle
%

\section{Introduction}

Galaxy clusters are important astrophysical laboratories and
cosmological probes (e.g. Sarazin 1986, Voit 2005, Borgani 2006,
Vikhlinin et al. 2009, Allen et al. 2011, B\"ohringer 2011).
While the latter references are based on X-ray observations of galaxy 
clusters, a lot of recent progress has also been made by optical
cluster surveys (e.g. Rozo et al. 2010) and millimeter wave surveys
using the Sunyaev-Zel'dovich effect (Reichardt et al. 2012, Benson et al. 2012, 
Marriage et al. 2011, Sehgal et al. 2011, PLANCK-Collaboration 2011). 
However, the currently most detailed view on the structure
and the properties of clusters comes from X-ray observations.
An X-ray survey is also the best means of efficiently
detecting galaxy clusters as gravitationally bound and well
evolved objects. X-ray observations thus provide  
statistically well defined, approximately mass selected
cluster samples, since (i) X-ray luminosity is tightly correlated to
mass (e.g. Pratt et al. 2009), (ii) bright
X-ray emission is only observed for evolved clusters with
deep gravitational potentials, (iii) the X-ray emission is
highly peaked and projection effects are minimized,
and (iv) for all these reasons the survey selection function
can be accurately modeled.

The X-ray emission originates in the several 10 Million
degree plasma trapped in the cluster's gravitational potential
well. In hydrostatic equilibrium, which is well approximated in
most clusters that are not just in a stage of collision and
merging, the intracluster density is tracing the equipotential 
surfaces. The observed X-ray luminosity is proportional to the
square of the plasma density with usually a very weak temperature
dependence in the X-ray energy band used by X-ray telescopes.
The X-ray image provides, therefore, very important
information on the mass distribution in the cluster and on
its structure, even though we can see only one projection of
the volume X-ray emission of the cluster 
(e.g. B\"ohringer et al. 2010). Systematic searches for
galaxy clusters in X-rays are thus the currently most established 
prerequisite for comprehensive astrophysical studies, as well 
as for cosmological model testing.

The {\sf ROSAT} All-Sky Survey (RASS, Tr\"umper 1993), is the only 
existing full sky survey conducted with
an imaging X-ray telescope, providing a sky atlas in which one can
search systematically for clusters in the nearby Universe.
So far the largest, high-quality sample of X-ray selected galaxy
clusters is provided by the {\sf REFLEX} Cluster Survey (B\"ohringer et al.
2001, 2004) based on the southern extragalatic sky of RASS at declination
$\le 2.5$ degree.  The quality of the sample has been demonstrated by showing
that it can provide reliable measures of the large-scale structure
(Collins et al. 2000, Schuecker et
al. 2001a, Kerscher et al. 2001), yielding cosmological parameters 
(Schuecker et al.  2003a, b; B\"ohringer 2011) in good agreement within the measurement 
uncertainties with the subsequently published
WMAP results (Komatsu et al. 2011).
The {\sf REFLEX} data have also been used to study the galaxy velocity dispersion -
X-ray luminosity relation (Ortiz-Gil et al., 2004), the statistics of
Minkowski functionals in the cluster distribution (Kerscher et al. 2001),
and to select statistically well defined subsamples like 
the HIFLUGCS (Reiprich \& B\"ohringer 2002) and REXCESS (B\"ohringer
et al. 2007). The latter is particularly important as a representative
sample of X-ray surveys to establish X-ray scaling relations (Croston et al. 2008,
Pratt et al. 2009, 2010, Arnaud et al. 2010) and the statistics of the morphological
distribution of galaxy clusters in X-rays (B\"ohringer et al. 2010).
We also constructed a catalog of superclusters from the latest version
of the REFLEX catalog comprising 164 superclusters including close pairs
of clusters (Chon et al. 2013).

Other galaxy cluster surveys based on the RASS
comprise  XBACs (X-ray-brightest Abell-type clusters; Ebeling et al. 1996), 
BCS ({\sf ROSAT} Brightest Cluster Sample) and eBCS
survey (Ebeling et al. 1997,1998),  RBS (RASS1 Bright Sample; De Grandi et al. 1999),  
NORAS (Northern {\sf ROSAT} ALL-SKY Survey Cluster Sample; B\"ohringer et al. 2000), 
the SGP survey (South Galactic Pole cluster survey; Cruddace et al. 2002),
MACS (Most Massive Galaxy Clusters; Ebeling et al. (2000) and the CIZA survey 
(Clusters in the Zone of Avoidance; Ebeling et al. 2002,
Kocevski et al. 2007). The RBS and SGP surveys were part of the early
efforts to create the {\sf REFLEX} survey. 

In this paper we describe the extension of the {\sf REFLEX} survey from the previous
flux limit of $3 \times 10^{-12}$ erg s$^{-1}$ cm$^{-2}$ in the 0.1 - 2.4 keV 
band to $1.8 \times 10^{-12}$ erg s$^{-1}$ cm$^{-2}$. The number of clusters
increases from 447 to 915 with this extension. We have already used this cluster
sample to assess the power spectrum of the galaxy cluster distribution (Balaguera-Antolinez 
et al. 2010) with the interesting finding that the bias behavior of clusters in 
two-point statistics is exactly what is predicted by the theoretical statistical models.
While this study was conducted when about 5\% of the galaxy cluster redshifts 
were still missing, we have now almost completed the spectroscopic follow-up 
observations in two observing campaigns in 2010 and 2011 
(Chon \& B\"ohringer 2012) which leaves only 7 galaxy clusters without redshift
information. This small number of missing redshifts will not significantly
affect the statistics of the survey described in the present paper and
in some first cosmological applications. Our plan for the publication of the
full cluster catalog in the near future will be based on the completion
of the redshift measurements. 
In the following we will use the term 
{\sf REFLEX II} for the extended cluster survey and {\sf REFLEX I} for the previous
cluster survey.

The paper is organized as follows. In chapter 2 we provide an overview on the 
global properties of the survey. Section 3 describes the determination of the
X-ray parameters of the clusters and section 4 provides an overview on the source 
identification process, the selection of cluster candidates and the description
of the follow-up observations. In section 5 we describe the construction of the
survey selection function and in section 6 we derive various statistical properties
of the {\sf REFLEX II} survey. In section 7 we compare our cluster detections 
to several other cluster surveys. Section 8 provides a discussion of the 
results and of the completeness and contamination of the {\sf REFLEX II} 
cluster sample. Section 9 comprises the summary and conclusions.

For the derivation of distance dependent parameters we use a geometrically flat
$\Lambda$-cosmological model with $\Omega_m = 0.3$ and $h_{70} = H_0/70$ km s$^{-1}$ 
Mpc$^{-1}$ = 1. All uncertainties without further specifications refer to 1$\sigma$
confidence limits.

\section{Survey Properties}

   \begin{table}
      \caption{Regions of the sky at the LMC and SMC excised from the Survey}
         \label{Tempx}
      \[
         \begin{array}{llllll}
            \hline
            \noalign{\smallskip}
 {\rm region}& {\rm RA} & {\rm range }  & {\rm DEC} & {\rm range}
  & {\rm area (ster)} \\
            \noalign{\smallskip}
            \hline
            \noalign{\smallskip}
{\rm LMC 1}  & 58  &\to  103^o  & -63 &\to  -77^o   & 0.0655 \\
{\rm LMC 2}  & 81  &\to  89^o  & -58 &\to  -63^o   & 0.0060 \\
{\rm LMC 3}  & 103 &\to  108^o  & -68 &\to  -74^o   & 0.0030 \\
{\rm SMC 1}  &358.5 &\to 20^o  & -67.5 &\to  -77^o & 0.0189 \\
{\rm SMC 2}  &356.5 &\to  358.5^o & -73 &\to  -77^o   & 0.0006 \\
{\rm SMC 3}  & 20  &\to   30^o   & -67.5 &\to  -72^o & 0.0047 \\
            \noalign{\smallskip}
            \hline
         \end{array}
      \]
\label{tab1}
   \end{table}
%

\begin{figure}
   \includegraphics[width=\columnwidth]{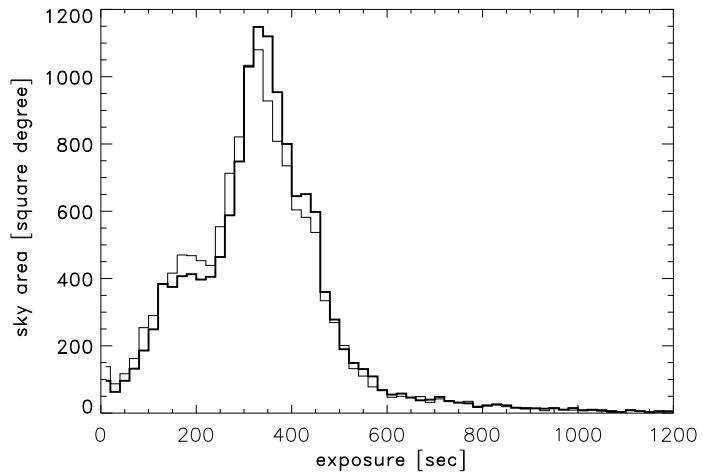}
\caption{Comparison of the exposure time distribution of the previous 
RASS II survey product used for {\sf REFLEX I} (thin line) and the improved 
survey analysis product RASS III (Voges et al. 1999) which is used for {\sf REFLEX II}. 
The increased total exposure coverage is mostly filling some gaps in the survey.
}\label{fig1}
\end{figure}

Like {\sf REFLEX I}, the extended survey covers the southern sky outside
the band of the Milky Way ($|b_{II}| \ge 20$ deg.) with regions around the Magellanic
clouds excised as defined in Table 1. The total survey area after
this excision amounts to 4.24 steradian (or 13924 deg$^{2}$) which
corresponds to 33.75 \% of the sky. Different from {\sf REFLEX I}, we 
use the refined RASS product RASS III (Voges et al. 1999) in which several 
small attitude errors were corrected and to which about 5 - 10\% survey 
exposure was added, which had been vetoed for RASS II mostly due to attitude 
problems  and for which a better attitude solution could be derived for RASS III.
The improvement of the survey exposure from RASS II to RASS III is illustrated
in Fig.~\ref{fig1}. This improvement allows us among other things to recover 
also some clusters above the {\sf REFLEX I} flux limit, which had too few 
counts to be detected in {\sf REFLEX I}.

The cluster candidate X-ray sources were selected from a reprocessed source list
of RASS III. The input source list for the reprocessing was taken from an intermediate
stage source detection analysis for the production of the public RASS source catalogue
(W. Voges, private communication) with a total of 157492 detections for 
the entire sky and 105968 detections at $|b_{II}| \ge 20$ deg.
The latter source list was reanalysed with the growth curve analysis (GCA) 
method (further described in section 3) to ensure that the 
flux of extended sources is well captured, 
since it is known that the standard analysis tuned for point sources 
tends to underestimate the flux of extended X-ray emission (B\"ohringer et al. 2000).
The results of the GCA processing were used to construct an
extragalactic source list of 5933 sources
with nominal flux \footnote{see section 3 for a definition of the nominal flux}
$\ge 1.8 \times 10^{-12}$ erg s$^{-1}$ cm$^{-2}$. For this source list double
detections of the same extended source were already removed. Since with the GCA method 
a stable plateau of the cumulative source count rate is not always found, depending
on the background properties or on source confusion, we also included a list of sources
with plateau flags 4 or 5 (indicating a non-stable plateau - for 
more details see B\"ohringer et al. 2000) comprising another 
1606 objects with fluxes $\ge 10^{-12}$ erg s$^{-1}$ cm$^{-2}$. 
For all these sources the cause of the non-stable plateau was inspected
and an appropriate correction (with a better background determination from a well
chosen comparison region) was applied individually. From both source
lists 4050 sources fall within the {\sf REFLEX} survey region. 
To make sure that we are not missing any X-ray source listed in the 
public RASS III source catalogue \footnote{the RASS source catalogs can 
be found at http://www.xray.mpe.mpg.de/rosat/survey/rass-bsc/ 
and http://www.xray.mpe.mpg.de/rosat/survey/rass-fsc/},
which was produced at a different survey analysis stage than
the extended preliminary list used above, 
we also cross-correlated our intermediate source list with this public data set and 
found 410 further sources above the flux limit to inspect. 
These sources do not necessarily 
represent sources missed in the earlier source list, since we use a recentering
algorithm in GCA and the GCA source positions do not directly correspond to the
detection positions of the input catalog sources
\footnote{This leads to the case that faint sources can be 
pulled into directly neighbouring bright sources and get lost. This is not a problem
for our cluster search since we are inspecting all sources visually anyway
noting any local complication.}. We find indeed a large number of double 
detections of very extended sources in this additional source list.
With this excercise we found only 8 previously not included clusters.
Since this comprises less than 1\% of the total sample we do not expect
any particular effect on the source detection statistics from this 
additional screening. 
Thus in total 4460 X-ray sources have been inspected in detail for the 
construction of the {\sf REFLEX II} cluster candidate list. The further selection
process of the candidates is described in section 4. 

\section{Determination of X-ray parameters} 

The X-ray count rates, fluxes, and luminosities
of the {\sf REFLEX} clusters are determined from the
count rate measurements provided by the GCA obtained in the energy
band defined by {\sf ROSAT} PSPC channels 52 to 201 
(approximately 0.5 to 2 keV) where the signal to noise is highest.
While we use the 0.5 to 2 keV band for the detection, we quote all
the fluxes and luminosities in the {\sf ROSAT} band, 0.1 to 2.4 keV. The 
conversion factor between both bands is almost constant over a wide
temperature range.
The GCA method is explained in
more detail in B\"ohringer et al. (2000, 2001) and we discuss only the 
relevant features of the technique here. 
The observed count rate is determined in two alternative 
ways. In one approach an outer radius of significant X-ray 
emission, $r_{sig}$, is determined
from the point where the increase in the $1\sigma$ error is larger
than the increase of the source signal. The integrated count rate
is then taken at this radius. In the second approach a horizontal
level is fitted to the outer region of the growth curve (at $r \ge r_{sig}$).
The value of this plateau is then adopted as the observed count rate. 
We use the second approach as the standard method but use 
also the first method as a check,
and a way to estimate systematic uncertainties in the count rate.
Specifically the error in the count rate is obtained from the square
root of the quadratic addition of the Poisson statistical error
of the count rate inside $r_{sig}$ and the difference of the
count rate determined by the two methods. There is no statistical 
justification of this error calculation, but it was adopted purely
as a practical measure of the combined uncertainty. The Poisson 
statistical error of the count rate within $r_{sig}$ also contains the 
uncertainty of the subtracted background count rate, which is a
minor contribution due to the fact that the region from which
the background surface brightness is estimated is much larger 
than the source region. For all further work we use the plateau
count rates. The radius at which the plateau is reached by the 
cumulative count rate curve is documented as the detection 
aperture radius, $r_{out}$.

Not in all cases we obtain a flat plateau, which can
be due to contaminating sources, very large source extent, or 
structure in the background. These cases, which comprise up to
10\% of all sources, are automatically flagged and the reason
for the distortion is carefully inspected and corrections are
performed individually. A conservatively estimated uncertainty 
is assigned to the source count rate in this process. 

To determine the cluster X-ray flux we convert the measured
count rate into an unabsorbed ``nominal'' X-ray flux for the 
{\sf ROSAT} band (0.1 to 2.4 keV), $F_n$, by assuming a thermal 
plasma spectrum for a temperature of 5 keV, a metallicity of
0.3 of the solar value (Anders \& Grevesse 1989), a redshift of zero,
and an interstellar hydrogen column density given for the line-of-sight
in the compilation by Dickey \& Lockman (1990),
as provided within EXSAS (Zimmermann et al. 1994).
The value of $F_n$ is used to make the flux cut independent of
any redshift information (since the redshift is not available
for all objects at the start of the survey).
With the redshift value at hand, the
unabsorbed X-ray flux, $F_x$,  is re-determined 
with an improved spectral model, where the temperature is now
estimated (iteratively) from the preliminarily derived X-ray luminosity
and the luminosity-temperature relation 

\begin{equation}
T = 3.31~ L_X^{0.332}~ h_{70}^{0.666} ~~,
\end{equation}

where $T$ is in keV and $L_X$ is measured in the 0.1 - 2.4 keV
energy band inside $r_{500}$~~ \footnote{$r_{500}$ is used for the
fiducial outer radius of the clusters, defined as the radius inside
which the mean mass density of the cluster is 500 times the critical
density of the Universe at the cluster redshift.}
in units of $10^{44}$ erg s$^{-1}$.
This relation is obtained from the $L_X$ - $T$ relation by
Pratt et al. (2009). Motivated by the results of
Reichert et al. (2011), we assume that the $L_X$ - $T$
relation has no redshift dependence. 

\begin{figure}
   \includegraphics[width=\columnwidth]{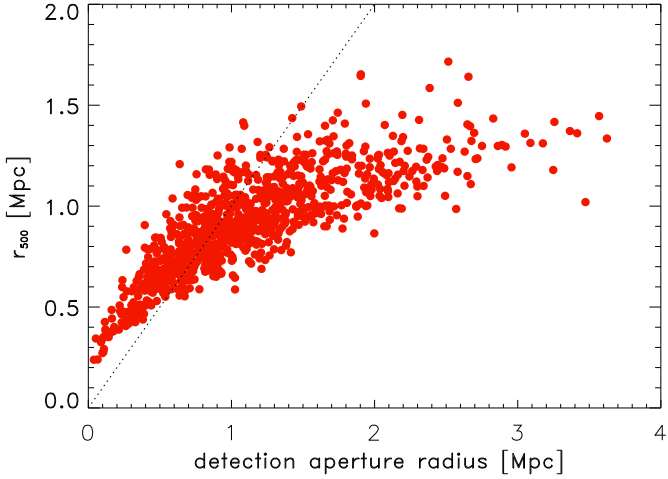}
\caption{Comparison of the radius of the detection aperture,
$r_{out}$, and the fiducial cluster radius, $r_{500}$, in physical
units. For less massive objects the surface brightness is lower and 
they are detected out to smaller scaled radii, while the objects with
the largest detection apertures are very luminous clusters in particular
those at high redshift where the PSF smearing increases the emission
region additionally. 
}\label{fig2}
\end{figure}

The estimated temperature of the cluster is now taken 
into account by folding the appropriate redshifted thermal 
spectrum with the instrument response and the 
interstellar absorption, leading to a revised flux, $F_x$ of 
the source (this correction is less than 5\% for sources with an
X-ray luminosity above $4 \times 10^{43}$ erg s$^{-1}$).
To obtain the cluster rest-frame luminosity
from the flux, $F_x$ we use the usual conversion with the cosmological
luminosity distance and further scale the luminosity by the ratio of
the luminosity integrated in the observed (redshifted) and rest frame
0.1 - 2.4 keV band. The latter is equivalent to the
K-correction of optical astronomy. 

\begin{figure}
   \includegraphics[width=\columnwidth]{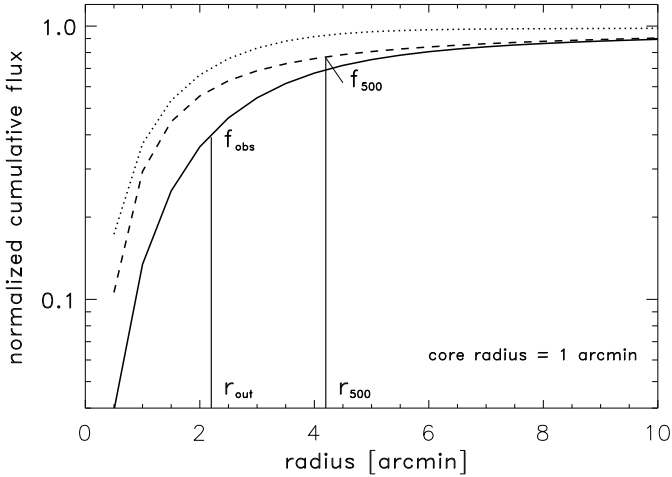}
\caption{Cumulative flux profiles for a point source with the 
average RASS PSF (dotted line), a $\beta$-model profile with 
a core radius of 1 arcmin (dashed line), and the same $\beta$-model 
profile convolved with the PSF (solid line). The profiles are 
normalized to unity at infinity. The observed flux fraction is read off from
the solid line at $r_{out}$ and the expected flux fraction at $r_{500}$ is 
marked as $f_{500}$. The ratio of the two values gives the correction factor
for the missing flux. 
}\label{fig3}
\end{figure}

The X-ray luminosities calculated and quoted in the following
(if not specified otherwise) are for the rest frame 0.1 to 2.4 keV
energy band. We aim for a determination of the X-ray luminosity
inside $r_{500}$. This is an improvement over the treatment in
the {\sf REFLEX I} publication, when precise scaling relation were not
available and we used a rather ad hoc scaling law.
The correction requires two considerations. We have to
estimate the cluster masses from $L_X$ to get $r_{500}$ and further
have to take into account, that the flux measurement is performed
with a limited angular resolution where the flux is spread outside the
$r_{500}$ aperture due to the point spread function (PSF) of
the survey. The half power radius of the RASS point spread function
is almost 1.5 arcmin, and thus this large PSF cannot be neglected.

$r_{500}$ is determined from the cluster mass ($M_{500}$) which
is estimated from the $M_{500}$ - $L_{X,500}$ relation. We use
a relation that is a compromise of the results of the studies by
Pratt et al. (2009), Vikhlinin et al. (2009), and Reichert et al.
(2011):

\begin{equation}
M_{500} = 2.48~  L_{X,500}^{0.62}~ E(z)^{-1}~ h_{70}^{0.242} ~~,
\end{equation}

and

\begin{equation}
r_{500} = 0.957~ L_{X,500}^{0.207}~  E(z)^{-1}~ h_{70}^{-0.586} ~~,
\end{equation}

where $M$ is in units of $10^{14}$ M$_{\odot}$, $r_{500}$
in units of Mpc, and $L_X$ is measured in the 0.1 - 2.4 keV
band in units of $10^{44}$ erg s$^{-1}$. 

In Fig.~\ref{fig2} we compare
$r_{500}$ with the detection aperture radius (both in
units of Mpc). For$\sim 57\%$ of the clusters the detection radius is 
smaller than $r_{500}$ and we have to correct for the missing flux.
The largest corrections are needed for the clusters with small
$r_{500}$, which are the low mass and low X-ray luminosity clusters.
The low mass systems have a lower surface brightness within
the self-similar model of cluster structure (e.g. Croston et al. 2008)
and thus their emission is not traced as far out as for massive
clusters. The largest aperture radii come from the most massive
clusters at higher redshifts where also the PSF smearing has an effect
and from some irregular, diffuse or double clusters.
For these clusters the flux will be corrected down by a few
percent.

To perform the missing flux correction we have to adopt a
generic model for the surface
brightness profile of the cluster. As in our previous studies and
many other studies in the literature we assume that clusters
can be described with good enough approximation 
in a self-similar fashion by a $\beta$-model
(Cavaliere \& Fusco-Femiano 1976) with a $\beta$-value of 2/3.
The observed cluster is then assumed to have
a model surface brightness profile folded with the point spread
function. Fig.~\ref{fig3} compares the cumulative flux profiles of a cluster
with a core radius of 1 arcmin, with and without folding with the PSF.
The curves are normalized by the flux integrated to infinity.
The recipe for correcting for the missing flux in this case is then
straightforward. The integrated flux at $r_{500}$, $F_{X,500}$,
is determined from the observed flux, $F_{obs}$, and the fractions,
$f_{obs}$, which can be read off from the PSF convolved curve
at the detection aperture radius ($r_{out}$) and $f_{500}$ the fraction
read off from the unconvolved curve at $r_{500}$. We then have

\begin{equation}
F_{X,500} = F_{obs} ~~ {f_{500} \over f_{obs} } ~~,
\end{equation}

and the missing flux fraction is

\begin{equation}
f_{miss}  =  {f_{500}  - f_{obs}  \over f_{500}} ~~.
\end{equation}

To implement this correction we have to know the core radius of the
cluster. 
An inspection of the self-similar intracluster electron
density profile of the clusters in the {\sf REXCESS} sample (Croston et al. 2008)
shows that the logarithmic slope of the density profile has
a value of one (that is at the core radius for $\beta$ = 2/3) at
about 0.15 to $0.2 \times r_{500}$, which implies $r_{500} \sim
5 -7 \times~ r_c$. The log-mean value of $r_{500} / r_c$ determined for
the clusters in the HIFLUGCS sample (Reiprich \& B\"ohringer 2002)
is about 9.5. We therefore adopted a value of  $r_{500} / r_c~ = 7$
but also explore values of 5 and 10. For the further calculations
we have then  tabulated the values of the curves shown 
in Fig.~\ref{fig3} for the full range of relevant core radii.

\begin{figure}
\begin{center}
   \includegraphics[width=7.5cm]{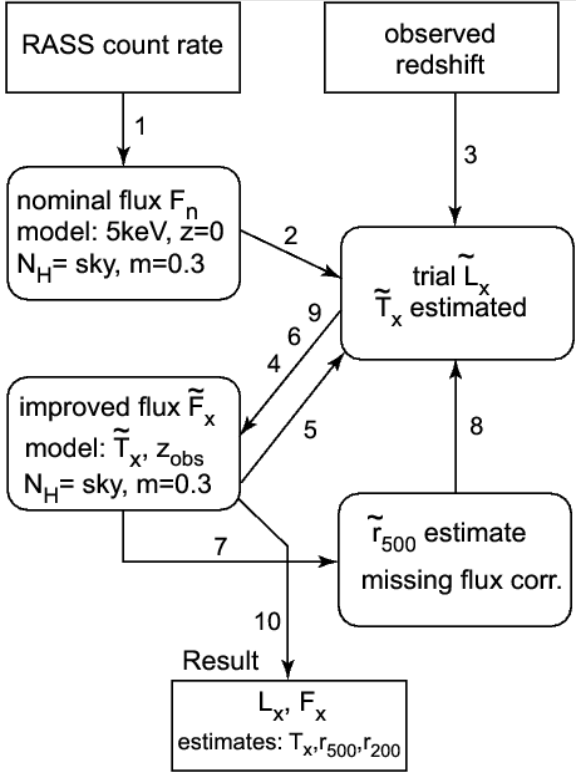}
\caption{Scheme of the iterative flux and luminosity correction taking
the temperature estimate, the K-correction and the missing flux correction
into account. In the first part the nominal flux, $F_n$, is calculated 
by assuming a temperature of 5 keV and a redshift of zero. In the second part,
in the loop comprising steps 4, 5, 6, an improved flux and luminosity is 
calculated taking the estimated temperature and measured redshift into account.
In the third part comprising steps 7, 8, 9 the missing flux is taken into
account and once more a better temperature estimate is included
in the new flux and luminosity calculation.
}\label{fig4}
\end{center}
\end{figure}

Note that different conventions have been used for the missing flux
correction. Ebeling et al. (1996, 1998) and De Grandi et
al. (1999) for example extrapolate their flux corrections to infinity. 
For the assumptions used here, with $r_{500} = 5, 7, 10 \times r_c$ the
fraction of the flux inside $r_{500}$ compared to infinity for a $\beta$-model
with slope 2/3 is 80\%, 84\%, and 90\%, respectively.
The agreement between the results of, for example, Ebeling et al. (1998) and 
B\"ohringer et al. (2000) indicates, however, a bias smaller
than this difference. We attribute this to the fact 
that our GCA method is capturing slightly more of the cluster flux 
than the other methods.

The two corrections, the first for the proper spectrum determined 
by temperature and redshift and the second for the missing flux, have
to be performed iteratively, since the $L_X - T$ and $L_X - r_{500}$
relations are defined for the corrected values. Fig.~\ref{fig4} shows
how this iterative calculation is performed in practice. In the loop of
steps 4, 5, and 6 the cluster parameters are stepwise improved
taking the estimated temperature and measured redshift into account.
A second loop with steps 7, 8, and 9 then includes the missing 
flux correction simultaneously iterating over the $L_X - T$ and
$L_X - r_{500}$ relations. In both loops the convergence is so
rapid, that two to three iterations are sufficient.

\begin{figure}
   \includegraphics[width=\columnwidth]{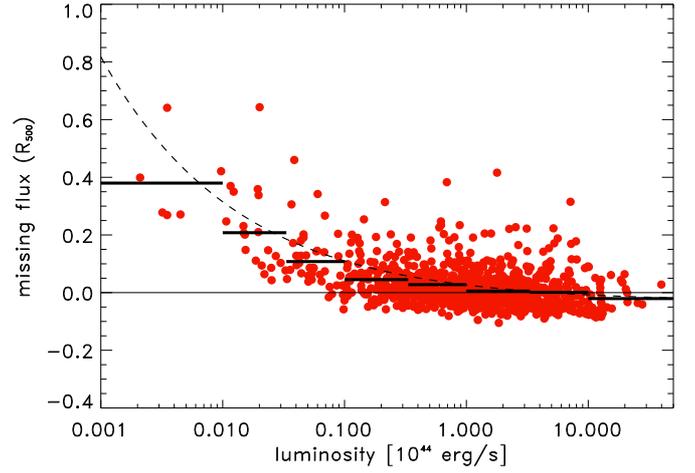}
\caption{Estimated missing flux as a function of X-ray luminosity for the
clusters in the {\sf REFLEX II} sample. The horizontal bars show the mean values
in 8 X-ray luminosity bins. The dashed line shows the best fitted power law
function with linear offset as given in Eq. 6.
}\label{fig5}
\end{figure}

We use the {\sf REFLEX II} results in the following to illustrate the relevance
of the missing flux correction. We see in Fig.~\ref{fig2} already that our detection
aperture captures most or all of the X-ray flux, $F_{X,500}$, for the majority
of the clusters and thus we expect the missing flux correction to be small,
except for the systems with the lowest X-ray luminosities. 
The resulting missing flux fraction that has to be corrected for
is shown in Fig.~\ref{fig5}. We found that one of the clearest and strongest 
dependencies of the missing flux parameter, $f_{miss}$, is on the X-ray 
luminosity. More X-ray luminous and more massive galaxy clusters have 
brighter X-ray surface brightness profiles and will therefore be traced out
to larger radii by the GCA method. This is by far the most important effect
that shows up in Fig.~\ref{fig5}. For clusters with luminosities above $10^{43}$ erg
s$^{-1}$, the mean correction is smaller than 2\%. The mean becomes even
slightly negative for clusters with $L_X > 10^{44}$ erg s$^{-1}$.
This figure also shows a fitted function of the mean of the correction
in the following form:

\begin{equation}
f_{miss} =  0.0575 \times L_{X~500}^{-0.389} -0.0309     ~~.
\end{equation} 

The most challenging objects for the flux correction
are small nearby groups with low surface
brightnesses. For the smallest objects the 
formal corrections can be 30 - 60\%. We should be careful, however, since
we are using scaling relations mainly calibrated for galaxy clusters and
the scaling and thus the corrections for these very low X-ray luminosities
marking the boundary of isolated giant ellitpicals and groups is uncertain. 

Thus for the bulk of the {\sf REFLEX II} clusters the corrections are small and
the adoption of the simple $\beta$-model for the description of all cluster
profiles, irrespective of their true structure, will not introduce 
a large error in the overall result. An inspection
of the results changing the $r_{500}$ - core radius relation to a ratio
of 5 or 10 introduces an additional uncertainty of the order of $5\%$.
Since this additional uncertainty is smaller than the overall flux
uncertainty, we do not include this uncertainty in the quoted
flux error.

A comparison of the cluster fluxes determined from the RASS
with those of deeper observations from e.g. the XMM-Newton or
Chandra observatories could in principle provide important tests
for the accuracy of the flux determination method. Our literature
survey in Reichert et al. (2011, Fig. C1) shows, however, that 
the flux determination from XMM-Newton observations of the same cluster
by different authors typically varies by factors of about 1.5, larger
than our flux uncertainties. A sensible comparison of REFLEX II fluxes
with deeper observations therefore requires a thorough and comprehensive 
new analysis of a large data set which is out of the scope of the 
present paper.

\begin{figure}
   \includegraphics[width=\columnwidth]{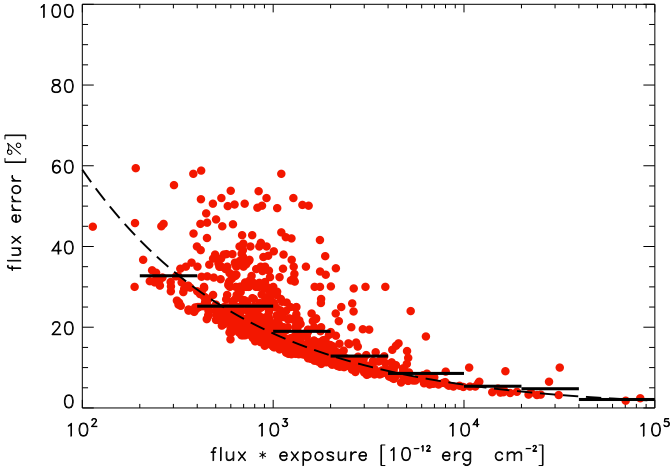}
\caption{Uncertainty of the flux determination as a function of the product
of flux and exposure time (a quantity close to the number of detected source
photons). The horizontal bars show the mean values of the uncertainty in 8
bins of flux $\times$ exposure. The dashed line shows the best power law fit
given in Eq. 7, which is close to a square root dependence on the number of
photons, as expected for Poisson statistical errors. 
}\label{fig6}
\end{figure}

Another important parameter, that has to be quantified for the modelling
of the cluster survey for future applications, is the error in the 
flux and luminosity determination. As explained above, the major source of
uncertainty in the flux (for most of the sources) is the Poisson statistics 
of the source photons. Thus in studying the dependence of the flux error
on various parameters, it comes as no surprise, that one of the clearest
dependencies found is that on the number of source photons detected. 
Using the number of source photons for the modelling of the survey would
imply the involvement of instrument specific operations to convert from
physical units into detected counts. Therefore, in looking for a very
similar, but more physical parameter that can be used as a substitute 
for the photon number,
we use the product of flux and exposure and find that the correlation with 
the flux error is indeed similarly good. 
This relation is shown in Fig.~\ref{fig6} where also a fit 
to this relation is presented.
The fit is indeed very close to a square root function,  
demonstrating that the Poisson-noise is the driving force for the error.
The error function shown can be expressed as:

\begin{equation} 
f_{err} = 603.6 \times \left( F_X~ \times {\rm exposure} \right)^{-0.505}  ~~,
\end{equation}

where $F_X$ is given in units of $10^{-12}$ erg s$^{-1}$ cm$^{-2}$.

The overall mean uncertainty is 20.6\%, significantly larger 
than for {\sf REFLEX I} due to the lower flux limit. Most of the data 
that appear in Fig.~\ref{fig6} as points up-scattered from the statistical 
line include those cases where sources have been deblended and new 
uncertainties including the deblending uncertainty have been calculated
case by case.

To allow the reader to easily reproduce our present results 
in connection with the {\sf REFLEX II} catalog to be published
we provide here all the conversion tables used
in the flux and luminosity determination  described above. 
The tables are the same as used for {\sf REFLEX I}. The printed version
only shows the first few lines of these tables, which are given 
in full size in the electronic version. Table 2 provides
the count rate to unabsorbed flux conversion for different temperatures
and interstellar column densities, while Table 3 shows the conversion
factors as a function of temperature and redshift. The conversion table
for the unabsorbed flux in the 0.1 to 2.4 keV to the flux in the
0.5 - 2 keV band (which is now most often used in the literature to
quote fluxes and luminosities of galaxy clusters) as well as the conversion
to bolometric flux can be found in the {\sf REFLEX I} catalog paper 
(B\"ohringer et al. 2004) \footnote{The conversion tables are also
available at http://www.mpe.mpg.de/$\sim$hxb/REFLEX/}.
To distinguish the {\sf REFLEX} cluster sources with their properties
and sky positions as determined with the described GCA method from 
the X-ray sources in the RASS catalog, we designate the {\sf REFLEX} clusters
with the name RXCJ+coordinate, as was done for the {\sf REFLEX I} catalog.

   \begin{table*}
      \caption{Count rate to flux conversion factors for different
        temperatures (for $Z = 0.3$ solar, $z = 0$) as a function of column density.
        The values quoted give the 0.1 - 2.4 keV flux per counts in the 0.5 to 2 keV band
         (channel 52 to 201) in units of $10^{-11}$ erg s$^{-1}$ cm$^{-2}$ counts$^{-1}$.
        The last column gives the hardness ratio for an assumed temperature of 5 keV,
         defined as  (counts(0.5 - 2 keV)-counts(0.1-0.4 keV))/(counts(0.5-2 keV)+counts(0.1-0.4 keV)).
         An extended version of this table is given in electronic form at CDS via anonymous ftp at
         cdsarc.u-strasbg.fr and our home page: http://www.mpe.mpg.de/$\sim$hxb/REFLEX.}
         \label{tab1}
      \[
         \begin{array}{llllllllllllll}
            \hline
            \noalign{\smallskip}
 N_H  & \multicolumn{12}{c}{\rm temperature} & {\rm ~~~~~HR} \\
10^{20} {\rm cm}^{-2}~~~ & 0.5  & 1.0 & 1.5 & 2.0 & 3.0 & 4.0 & 5.0 & 6.0 & 7.0 & 8.0 & 9.0 & 10.0 
{\rm ~keV} &   \\
            \noalign{\smallskip}
            \noalign{\smallskip}
            \hline
            \noalign{\smallskip}
  0.10 & 1.281& 1.413& 1.751& 1.831& 1.868& 1.880& 1.887& 1.893& 1.897& 1.900& 1.901& 1.902&0.003\\
  0.30 & 1.291& 1.422& 1.761& 1.842& 1.879& 1.891& 1.898& 1.904& 1.908& 1.910& 1.912& 1.913&0.089\\
  1.00 & 1.325& 1.453& 1.796& 1.880& 1.917& 1.929& 1.936& 1.942& 1.946& 1.948& 1.950& 1.951&0.323\\
  3.02 & 1.429& 1.543& 1.900& 1.989& 2.028& 2.040& 2.046& 2.052& 2.055& 2.057& 2.059& 2.059&0.691\\
 10.00 & 1.840& 1.886& 2.278& 2.383& 2.427& 2.438& 2.442& 2.446& 2.449& 2.450& 2.451& 2.450&0.943\\
 30.20 & 3.654& 3.213& 3.583& 3.718& 3.766& 3.766& 3.762& 3.760& 3.756& 3.753& 3.749& 3.744&0.978\\

          \noalign{\smallskip}
            \hline
         \end{array}
      \]
   \end{table*}
%
%
   \begin{table*}
      \caption{K-correction table for different temperatures and redshifts. The given value is to be
      multiplied with the luminosity in the observed band to obtain the luminosity in the rest frame band.
      An extended version of this table is given in electronic form at CDS via anonymous ftp at 
          cdsarc.u-strasbg.fr and our home page: http://www.mpe.mpg.de/$\sim$hxb/REFLEX.} 
         \label{tab2}
      \[
         \begin{array}{lllllllllllll}
            \hline
            \noalign{\smallskip}
  {\rm redshift~~~} & \multicolumn{12}{c}{\rm temperature (keV)} \\
 & 0.5  & 1.0 & 1.5 & 2.0 & 3.0 & 4.0 & 5.0 & 6.0 & 7.0 & 8.0 & 9.0 & 10.0 \\
            \noalign{\smallskip}
            \hline
            \noalign{\smallskip}
0.05 & 1.0026 & 0.9935 & 0.9867 & 0.9838 & 0.9800 & 0.9771 & 0.9750 & 0.9733 & 0.9720 & 0.9709 & 0.9700 & 0.9693\\
0.10 & 1.0086 & 1.0253 & 0.9852 & 0.9700 & 0.9596 & 0.9540 & 0.9502 & 0.9472 & 0.9449 & 0.9431 & 0.9415 & 0.9402\\
0.15 & 1.0126 & 1.0258 & 0.9806 & 0.9611 & 0.9450 & 0.9359 & 0.9299 & 0.9253 & 0.9217 & 0.9189 & 0.9166 & 0.9147\\
0.20 & 1.0273 & 1.0416 & 0.9771 & 0.9528 & 0.9314 & 0.9192 & 0.9112 & 0.9050 & 0.9003 & 0.8966 & 0.8936 & 0.8911\\
0.25 & 1.0452 & 1.0799 & 0.9880 & 0.9489 & 0.9197 & 0.9041 & 0.8940 & 0.8864 & 0.8806 & 0.8760 & 0.8724 & 0.8693\\
0.30 & 1.0497 & 1.0820 & 0.9833 & 0.9401 & 0.9070 & 0.8891 & 0.8775 & 0.8686 & 0.8619 & 0.8566 & 0.8523 & 0.8488\\
0.40 & 1.0584 & 1.0850 & 0.9768 & 0.9254 & 0.8837 & 0.8614 & 0.8469 & 0.8359 & 0.8276 & 0.8211 & 0.8159 & 0.8115\\
            \noalign{\smallskip}
            \hline
         \end{array}
      \]
   \end{table*}

Two further important parameters delivered by the GCA method
which help to characterize the X-ray sources are the 
spectral hardness ratio and the source extent.
The hardness ratio, $HR$, is defined as
$HR = {H - S \over H + S}$
where $H$ is the hard and $S$ the soft band source count rate
(both determined for the same outer aperture radius). The hard 
and soft bands are defined to comprise the {\sf ROSAT} PSPC energy channels
from 10 to 40 and from 52 to 201, which correspond roughly to the energy
bands 0.1 - 0.4 keV and 0.5 to 2.0 keV, respectively. The hardness ratio
expected for galaxy clusters with temperatures above 2 keV depends mostly
on the interstellar hydrogen column density. Since this parameter is known
we can approximately calculate the expected hardness ratio and use this
parameter as one criterion in the identification.

The source extent is another important piece of information for identification
purposes, since the majority of the extended sources are expected to be 
clusters of galaxies. This is assessed in two ways.
In the first analysis a $\beta$-model profile (Cavaliere \& Fusco-Femiano 1976)
convolved with the averaged survey point spread function (PSF)
is fitted to the differential
count rate profile (using a fixed value of $\beta$ of $2/3$) yielding
a core radius, $r_c$, and a fitted total count rate. Secondly, a
Kolmogorov-Smirnov (KS) test is used to estimate the probability that
the source is consistent with a point source. The source is flagged to
be extended when the KS probability is less than
0.01. Tests with X-ray sources which have been identified with
stars or AGN show a false classification rate as extended sources
of typically less than 5\%. We make mostly use of the KS test 
results for the source identification.

\section{Source identification and selection of the cluster candidates}

Further selection of cluster candidates from the flux limited list of X-ray
sources was based on the experience acquired with {\sf REFLEX I}. We cannot rely 
on the detection of an extent of the X-ray emission for this selection, since a fraction
of the clusters at higher redshifts will not be resolved in the survey with its 
broad PSF of $\sim 1.5$ arcmin. We therefore have to consider all X-ray sources. 
The construction of {\sf REFLEX I} was partly based on a statistical characterization
of the X-ray sources through the possible coincidences with 
galaxy over-densities in the COSMOS Survey of the Royal Observatory 
Edinbourgh \footnote{The COSMOS Survey is the results of digital
scans of photographic plates taken at the UK Schmidt telescope in Australia, see
http://www-wfau.roe.ac.uk/sss/}. The depth of this photographic survey is not
sufficient to make a unique decision, however, in particular for the most distant
clusters in the sample. Therefore already for the identification
of the possible cluster candidates in {\sf REFLEX I} we had to rely on the 
combination of all other available information.

The main sources of information considered here for the selection are: 
the detailed X-ray properties determined from
the RASS, the large number of identifications of X-ray sources obtained from 
NED\footnote{see http://ned.ipac.caltech.edu/}, and digital sky images
obtained from DSS\footnote{see http://archive.stsci.edu/dss/}. An important
aspect of the inspection of DSS images included  their combination
with X-ray surface brightness overlays.

\begin{figure}
\begin{center}
   \includegraphics[width=7.5cm]{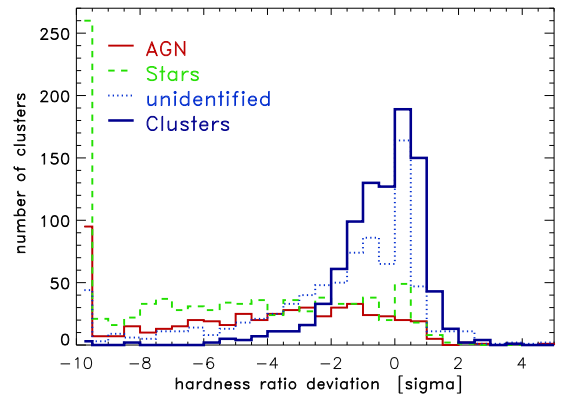}
\caption{Hardness ratio parameter distribution of various groups of 
X-ray source classes: galaxy clusters, AGN, stars, and
unidentified objects. The hardness ratio parameter
shown is the deviation in units of sigma of the observed hardness ratio
value from the expectation for a 5 keV cluster at $z = 0$ with given interstellar
absorption in the line-of-sight.
}\label{fig7}
\end{center}
\end{figure}

\begin{figure}
\begin{center}
   \includegraphics[width=7.5cm]{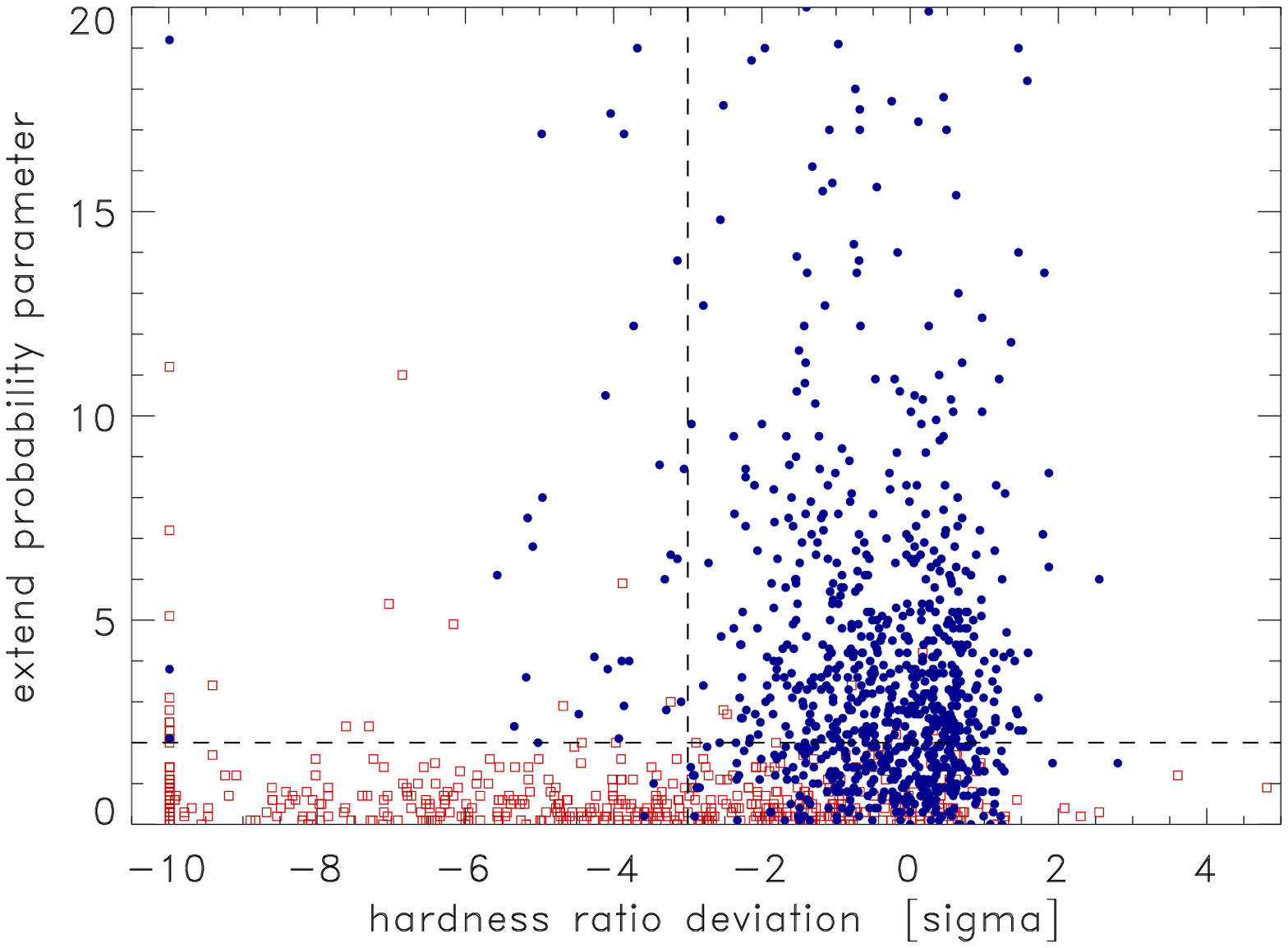}
   \includegraphics[width=7.5cm]{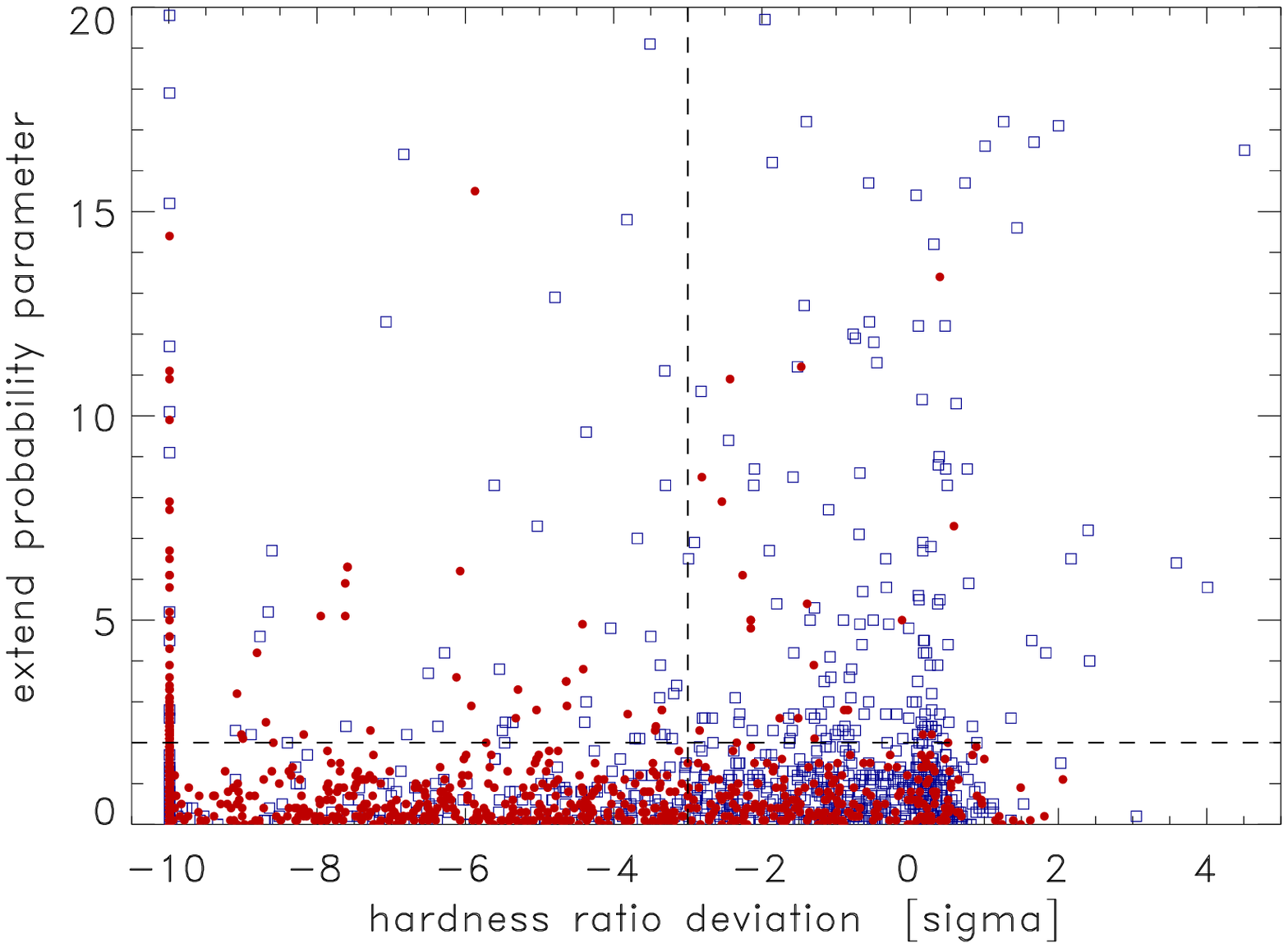}
\caption{Comparison of the distribution of the hardness ratio
and source extent parameters of X-ray sources
identified  as clusters (blue dots), and tentatively classified AGN (red squares)
in the upper panel and stars (red dots) as well as unidentified sources
(blue squares) in the lower panel.
}\label{fig8}
\end{center}
\end{figure}

The way we make use of the X-ray spectral and source extent information 
in the selection process is explained in the following. To calculate
the expected hardness ratio of a cluster, we use the same model assumptions
as for the nominal flux: a temperature of 5 keV, a metallicity of 0.3 solar
and redshift $z=0$. The interstellar absorption used is the one for the sky
position of the cluster. We compare the expected HR values to the observed ones
with their estimated errors and determine the deviation of the two results
in terms of the sigma errors. Fig.~\ref{fig7} compares the hardness ratio deviations
for different X-ray source types. Cluster sources pile up on the high end
of the hardness ratio and the most important discrimination is that
towards softer sources. A conservative value to mark an X-ray source to
be too soft for a cluster is therefore taken as a value of 
less than $-3\sigma$. We see in Figs.~\ref{fig7} 
and \ref{fig8} a few cluster sources below this
threshold. They are on one hand very bright nearby clusters where the very
good photon statistics causes a deviation from the simple fiducial model.
There is no doubt about their identification as clusters, however. Another
fraction of these sources are those which are contaminated by soft 
point sources which have been deblended as well as possible 
(we did not re-estimate the hardness ratio after deblening since this would
be more uncertain than a simple flux deblending).

We can see from Fig.~\ref{fig7} that most of the stars are softer than clusters, 
while for the AGN only a fraction of the X-ray counterparts can be ruled out as
too soft. Also for the sample of unidentified sources only a smaller fraction
can easily be ruled out from being clusters based on their hardness ratio
parameter.

For the source extent evaluation we use the KS probability parameter with
a value of less than 0.01 of being a point source to characterize
highly likely extended sources. Tests with samples of known point sources
(e.g. AGN) show that the probability of these sources
being characterized as extended is less than 5\%. Several factors, like
an uneven survey PSF, influence of background noise, and slight attitude 
shifts in those orbits that contribute to the photon counts of the 
source can contribute to produce an artificial extent which is not 
accounted for in the KS test. This explains why the fraction of false
extents is larger than predicted by the KS test. Nevertheless, 
the fact that a cut at 1\% results in a
contamination of clearly less than 5\% is a very reassuring
result that the spatial characterization of the X-ray sources works
very well.

Fig.~\ref{fig8} shows the combined distribution of the hardness ratio and extent
parameter for X-ray sources identified as galaxy clusters, tentatively
identified with stars and AGN, and unidentified sources. A large fraction
of the clusters occupy the upper right corner of extended, hard sources
which is less frequented by the other sources. Thus the majority of the
{\sf REFLEX II} cluster sources are quite readily identified, in particular as 
most of these clusters show up as optical galaxy concentrations
in DSS sky images. However, 31\% of the cluster sources have a point source
probability larger than 0.01 and are classified as pointlike in our conservative
scheme. Therefore the large effort of inspecting all the sources has to
be conducted to arrive at the desired high sample completeness.

\begin{figure}
\begin{center}
   \includegraphics[width=7.5cm]{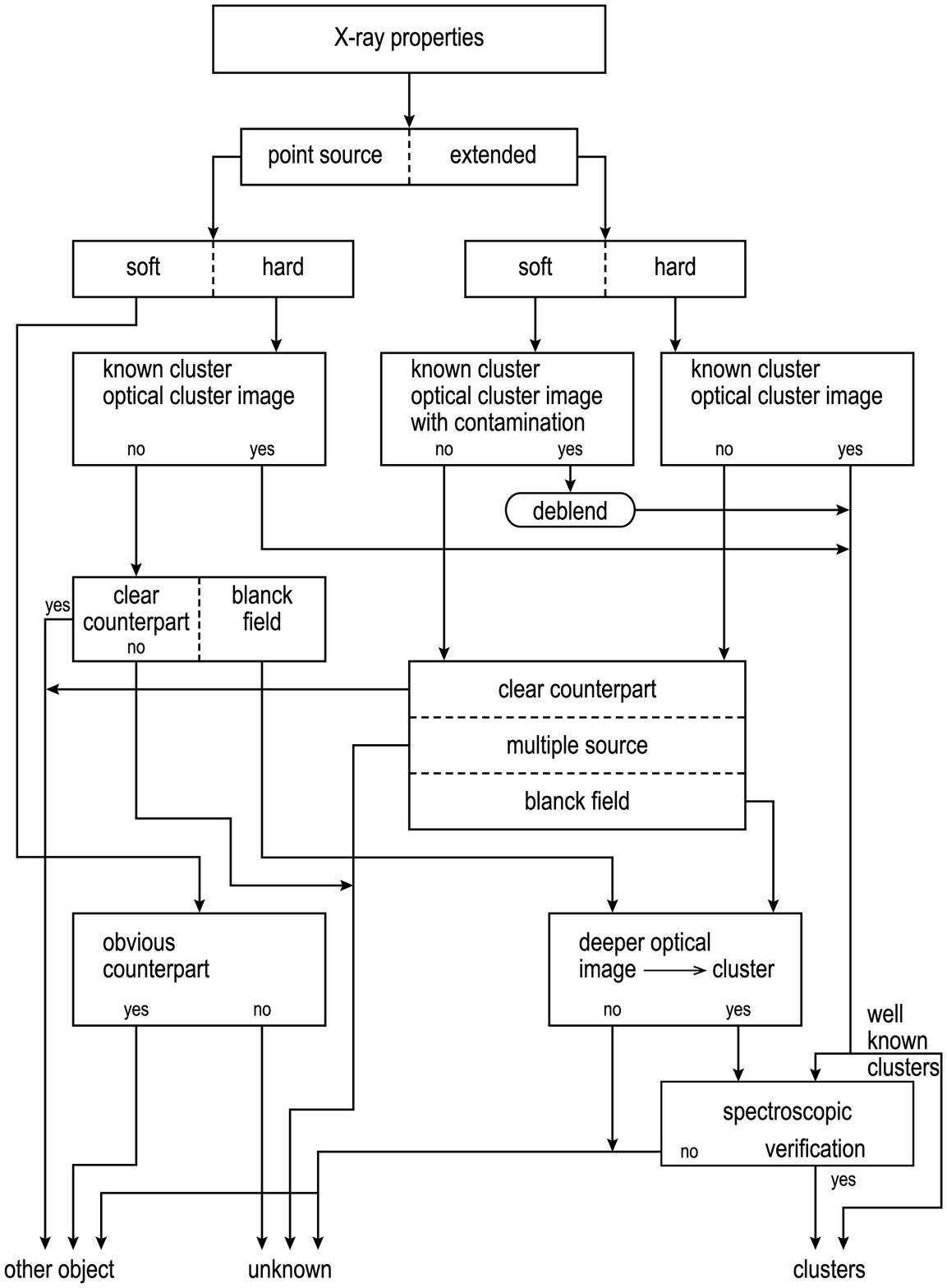}
\caption{Scheme of the galaxy cluster selection process from
the list of X-ray sources. The list starts at the top with
the classification of the sources by means of the
X-ray parameters for source extent
and hardness ratio. For the further identification this 
information is combined with information from NED, the literature 
and available sky images.
}\label{fig9}
\end{center}
\end{figure}

Fig.~\ref{fig9} gives an overview on the cluster candidate selection strategy. For
X-ray sources already clearly identified in the literature we adopt this
classification if we have no doubt about it after an inspection of the
observational information. To conservatively rule out further X-ray 
sources from being a cluster we typically use at
least two negative criteria: (i) the source is too soft and is consistent with
being a point source, (ii) the source is a point source and coincident with
a bright star, known galactic X-ray source, or AGN, (iii) the source is flagged as 
extended and coincident with a nearby galaxy, galactic HII region, or supernova
remnant, (iv) the source is flagged as extended but best explained by multiple sources
with a hardness ratio or other properties inconsistent with being a cluster. For a number 
of point sources with spectral properties not inconsistent with a cluster
we are left with no classification. In case there is any weak indication of
some faint galaxies at the X-ray center or for completely blank fields
we have taken deeper images as part of our observation runs. 
Inspecting more than 40 such borderline cases that had not been flagged as 
promising candidates we did not find a cluster. Since most promising fields 
have been targeted, we are convinced
that we have reached a high completeness in our cluster identification of
the flux limited source list. 
On the positive selection side we could be more generous to include weak
cluster candidates since with the follow-up spectroscopic identification, 
which is described below, the false clusters are revealed. 

In some cases we found that the clusters were contaminated by point sources. 
In this case we did the best effort to deblend the point source from the cluster 
emission and added the estimated uncertainty of the deblending procedure to 
the flux error. Specifically, in addition to the visual inspection we used the 
procedure described in B\"ohringer et al. (2000) testing for irregularities in the 
statistics of photon counts in annuli sectors of the source to flag potential
off-center contaminating sources. While this method also provides a first 
estimate of the possible contaminating flux, it does not provide a distinction
between cluster substructure and source contamination. After a further test
to decide if the contaminating source is most likely a point 
source or not, we decided to deblend the
clearly pointlike sources while accepting the flux of non-pointlike sources as 
substructure. In special cases the contamination by a non-cluster source 
becomes evident from the difference between the hardness ratio of the 
contaminating source and the cluster emission. Fig. 3 in Chon
\& B\"ohringer (2012) gives a nice example for such infrequent cases. In another 
class of X-ray sources we could either see a galaxy grouping or cluster at
the X-ray position or we found information on the previous identification 
of a group or cluster, but it was obvious that the X-ray emission is dominated
most likely by an AGN and the possible cluster emission is below the flux 
threshold of the survey. In these cases we identified the source as
AGN in a cluster and we will provide information on such sources with the 
publication of the cluster catalog, as we have done in the catalog publication
for {\sf REFLEX I} (B\"ohringer et al. 2004).

In total we arrive very roughly at the following tentative classification
for the non-cluster sources: 46\% stars, 30\% AGN, 3\% galaxies and galactic
sources, 21\% unidentified sources. It shows that at this flux
level a large number of sources can be readily identified with high probability
and the number of unidentified sources is a small, but far from a negligible
fraction.

The spectroscopic confirmation and redshift measurement by follow-up 
observations has been described in detail for {\sf REFLEX I} in Guzzo et al. 
(2009) and more information on the {\sf REFLEX II} follow-up is given in
Chon and B\"ohringer (2012). In summary almost all the dedicated 
follow-up observations for {\sf REFLEX II} 
have been carried out at the 3.6m and NTT telescopes
at La Silla between 2000 and 2011 using the EFOSC instruments. In most 
cases multi-slit spectroscopy was used providing spectra for typically 
about 7 cluster members. To increase the efficiency of the observing runs
long slit observations with two or three targets per slit were performed
in particular for nearby clusters where the inclusion of a clear cluster
BCG helped to get a unique redshift. The typical galaxy redshift accuracy
reached in the follow-up observations is 50 - 60 km/s.

The complete identification process provided a catalog of currently 915 
galaxy clusters. This sample does not include a predefined lower 
limit on the total source photon number.

\begin{figure*}
\centering
\includegraphics[width=17.0cm]{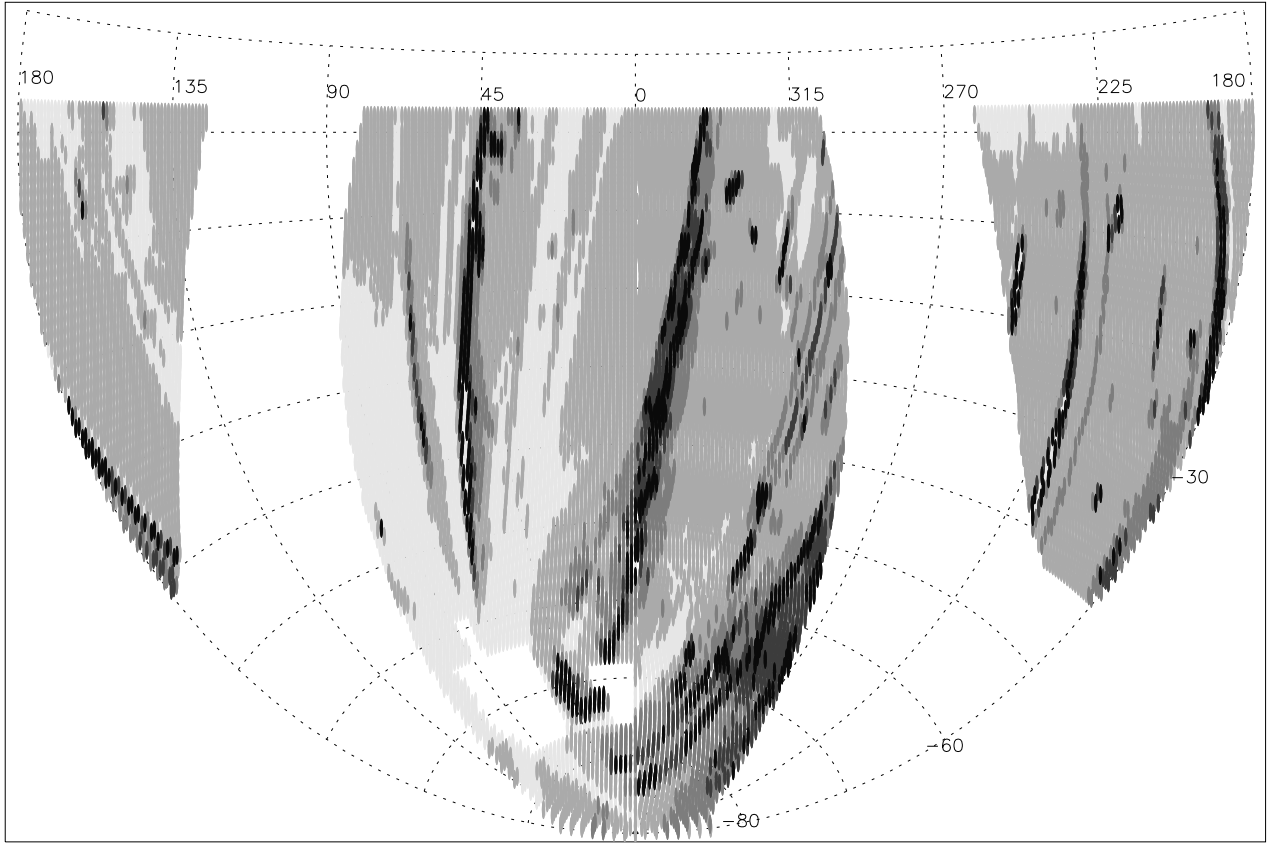}
\caption{Sensitivity map of RASS III in the area of the {\sf REFLEX II} survey.
Five levels of increasing grey scale have been
used for the coding the sensitivity levels given in units of the number of
photons detected at the flux limit:
$ > 60$ , $30 - 60$ , $20 - 30$, $15 - 20$, and $< 15$,
respectively. The coordinate system is equatorial for the epoch
J2000.
}
\label{fig10}
\end{figure*}

\begin{figure}
\includegraphics[width=\columnwidth]{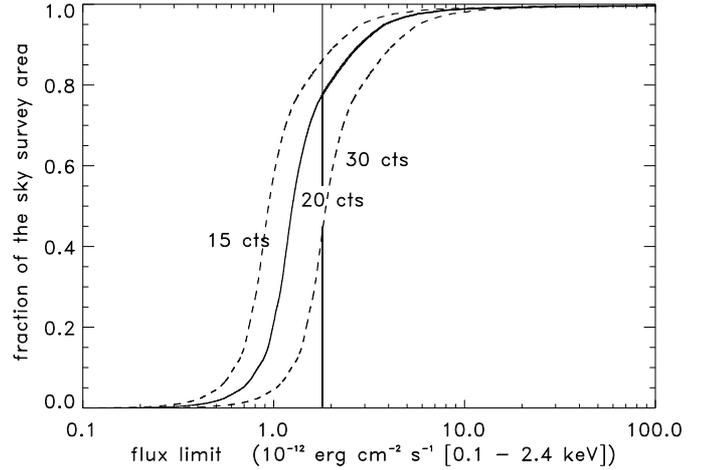}
\caption{Effective sky coverage
of the {\sf REFLEX II} sample. The thick line gives the effective sky area
for the nominal flux limit of $1.8 \cdot 10^{-12}$ erg s$^{-1}$ cm$^{-2}$
and a minimum number of 20 photons per source (as used e.g. for the correction
of the $\log$N$-\log$S-curve shown in Fig.~\ref{fig15}. For further details
see text).
}\label{fig11}
\end{figure}

\begin{figure}
\includegraphics[width=\columnwidth]{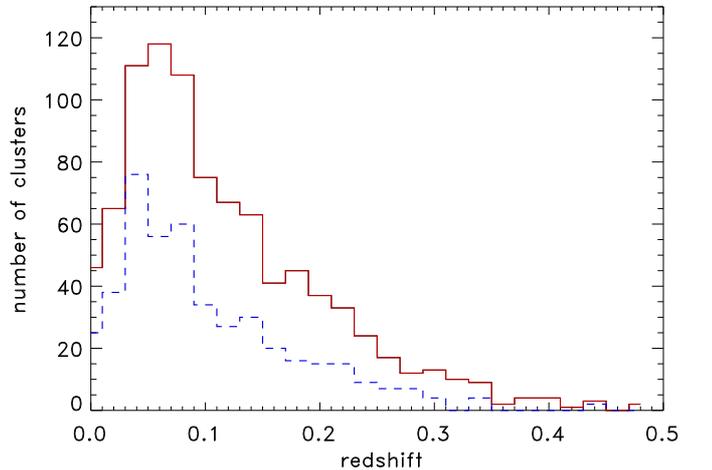}
\caption{Redshift distribution of the clusters in the {\sf REFLEX II}
(solid line) compared to the {\sf REFLEX I} sample (dashed line).
The most distant cluster at $z = 0.537$ is not represented in the plot.}
\label{fig12}
\end{figure}

\begin{figure}
\includegraphics[width=\columnwidth]{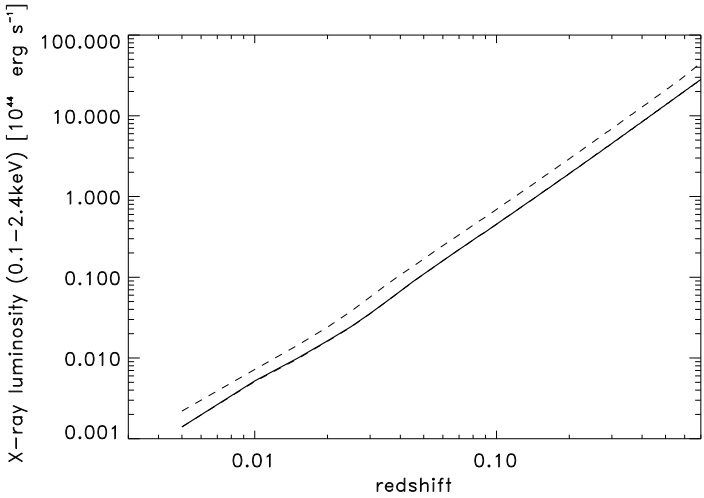}
\caption{X-ray luminosity limit for the cluster X-ray source detection
as a function of redshift. The solid line shows the median for the survey area
while the dashed line shows the values for the 10\% area with the lowest sensitivity.
If we would plot other percentiles from 30 to 90\%, they would hardly be 
distinguishable from the median line.}
\label{fig13}
\end{figure}

\begin{figure}
\includegraphics[width=\columnwidth]{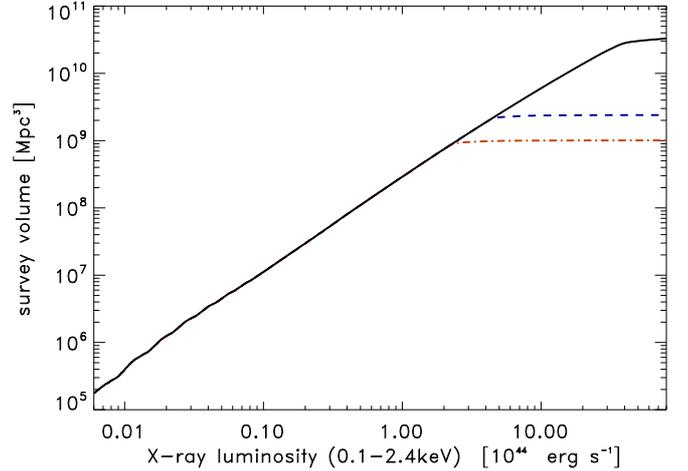}
\caption{Effective survey volume probed by {\sf REFLEX II} as a function of X-ray
luminosity. The solid curve shows the estimated survey volume for a
redshift limit of $z = 0.8$, while the dashed lines are determined for a
redshift limit of $z = 0.3$ (blue, dashed) and $z= 0.22$ (red, dashed-dotted), respectively.
}
\label{fig14}
\end{figure}

\section{Survey selection function}

For the analysis and modelling of the {\sf REFLEX II} cluster data we need to 
know the survey selection function. In a first step we calculate the nominal 
flux limit for each position in the {\sf REFLEX II} survey area (in the same way as was 
done for {\sf REFLEX I} in B\"ohringer et al. 2004). For this we use the exposure time 
of RASS III and the interstellar column density taken from Dikey \& Lockmann
(1990). Using the same count rate to flux conversion calculation as for the
determination of the nominal flux above, that is assuming a temperature of 5 keV,
a redshift of zero, and a metallicity of 0.3 solar, we calculate the flux 
corresponding to one source photon for the given RASS III exposure time.
We call this the sensitivity map of the survey and show it in Fig.~\ref{fig10}.
In the figure this map is show in the form of number of detectable photons for
the nominal flux limit.
Large stripes of low sensitivity are due to low exposure areas, that result from
times when the detectors on {\sf ROSAT} were switched off during the high 
particle flux in the South Atlantic Anomaly region. 
Compared to the corresponding map for {\sf REFLEX I} in B\"ohringer et al. (2004)
several of the smaller low sensitivity regions disappeared, due to the 
more comprehensive attitude solution of the RASS III survey product.
The sensitivity map is defined on a 1 deg$^2$ sky pixel grid. This is good
enough for our purpose, since the exposure variations are not large over these 
angular scales, given that the {\sf ROSAT} PSPC camera has a 2 degree diameter 
field-of-view.

To calculate the nominal flux limit for the survey as a function 
of sky position, we have to impose a limit for the minimum number of counts 
for the acceptable detection of an X-ray source. This value may
not be fixed, but different values can be used for different applications 
depending on the need of accuracy versus large number of objects. 
For example, in the modeling of {\sf REFLEX I} a 30 photon limit 
was for used to determine the X-ray luminosity function,
while for the determination of the density fluctuation power spectrum 
we preferred to maximize the number of data points in space and used 
a limit of 20 source photons. Going deeper in flux with {\sf REFLEX II}, 
the preferred number will be a count limit of 20 source photons. 
The motivation for this choice will be given below.

Fig.~\ref{fig11} shows the resulting cumulative sky area covered by the 
survey as a function of the nominal flux limit for three different 
detection count limits, 15, 20 and 30 counts. In addition to this 
variable count limits we have the fixed flux cut at $1.8 \times 10^{-12}$ erg s$^{-1}$
cm$^{-2}$, shown as a vertical line in the Figure, which is taken 
as the minimum flux limit in all cases. We note that for the 
three cases, 86.1\%, 77.6\%, and 43.8\% of the sky area are covered 
by the minimum flux limit, respectively, and only the smaller 
remaining parts have a higher limiting flux.  

Note that the effective sky coverage function is based on the 
assumption of a 100\% detection efficiency above the flux limit. 
This is justified, since our flux limit is well 
above the detection limit which has a soft boundary between $1$ and 
$5 \times 10^{-13}$ erg s$^{-1}$ cm$^{-2}$ (see Voges et al. 1999, e.g. Fig. 6).
Even though this detection limit mostly represents the bulk of the sources
which are pointlike, our flux limit is still high enough to 
also provide a nearly unit detection probability for the peaked 
extended sources of clusters of galaxies.
No flux errors have been folded into this function. The error folding
is performed in our modeling at a subsequent stage.

For the determination of the X-ray luminosity function and other cosmological
statistics like the power spectrum (Balaguera-Antolinez et al. 2010), we need to know
the selection of the clusters according to their true luminosity. We calculate 
this in the form of the limiting luminosity that will be detected in the survey
as a function of sky position and redshift. We call this multi-dimensional function
the luminosity selection mask of the {\sf REFLEX II} survey. This mask was for example 
used to construct mock samples from cosmological N-body simulations in the 
study of the {\sf REFLEX II} power spectrum (Balaguera-Antolinez et al. 2010).

\begin{figure}
\includegraphics[width=\columnwidth]{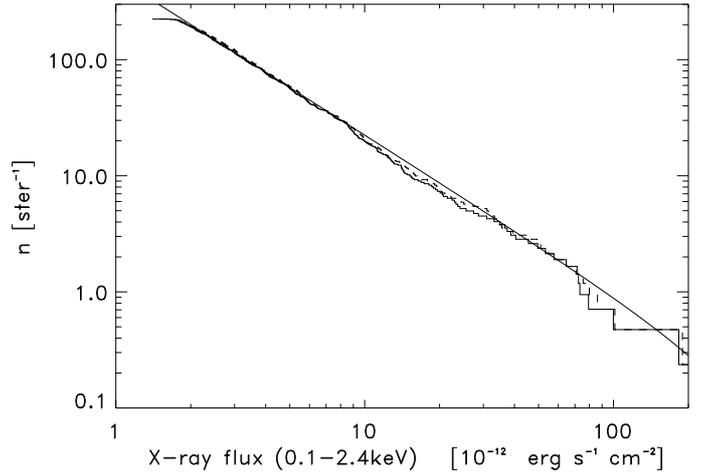}
\caption{LogN-logS distribution of the {\sf REFLEX II} clusters. The solid step
function shows the logN-logS-function
for the observed flux, $F_X$, and the dashed function the flux corrected to an aperture
of $r_{500}$. The solid line shows a power law function convolved
with a flux dependent flux error and fitted to the data by means of a
maximum likelihood method. The underlying power law function has a slope of
$1.36 (\pm 0.07)$.}
\label{fig15}
\end{figure}

\begin{figure}
\includegraphics[width=\columnwidth]{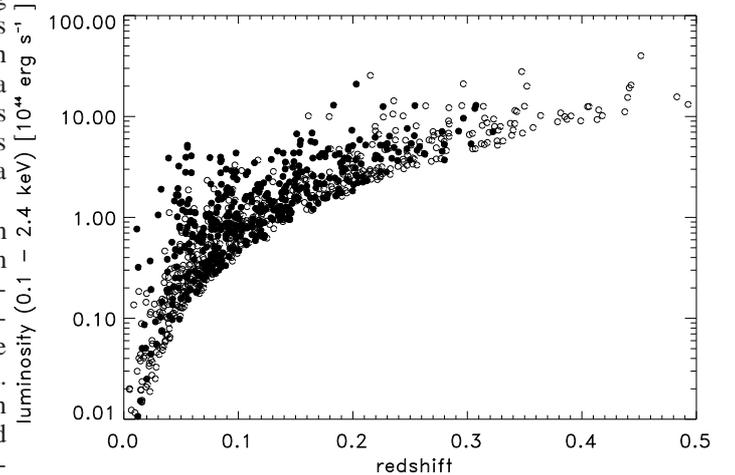}
\caption{$L_X$ and redshift distribution of the {\sf REFLEX II} clusters. The 
clusters identified by those listed by Abell (1958) and Abell et al. (1989) are
shown with filled dots, the non-Abell clusters by open symbols.
}\label{fig16}
\end{figure}

\begin{figure}
\includegraphics[width=\columnwidth]{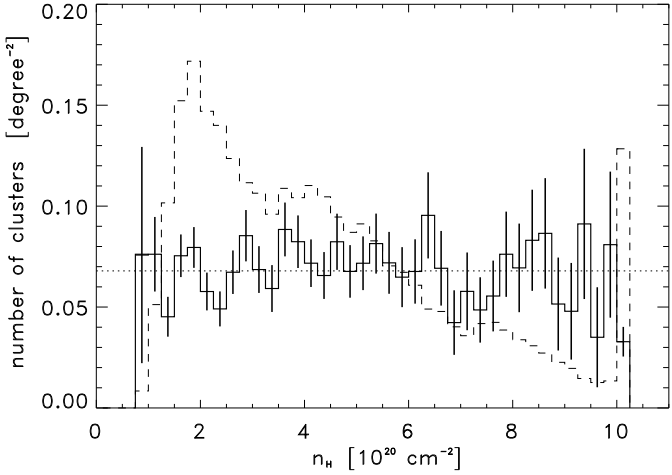}
\caption{Surface density of {\sf REFLEX II} clusters in the sky as a function
of the interstellar hydrogen column density, $n_H$. The dashed line shows the distribution
of $n_H$-values in the {\sf REFLEX} region for comparison. The dotted line shows the
mean value of 0.068 degree$^{-2}$. Note that the regions with a flux limit
smaller than the nominal flux limit have been scaled to a smaller effective
area as explained in the text. The last bin contains all regions with
$n_H > 10^{21}$ cm$^{-2}$.  
}\label{fig17}
\end{figure}

The selection mask is determined in the following way. For each sky pixel of
1 deg$^2$ size we calculate from the nominal limiting flux for a given redshift
the limiting luminosity (first for 5 keV and unredshifted spectrum) while iterating 
over the correction for the estimated temperature and redshifted spectrum.
We also correct in this iterative procedure for the missing flux according to 
Equation (6). The mask is tabulated for 160 redshifts out to $z = 0.8$ and for
13951 sky pixels. The mask will be published together with the {\sf REFLEX II} cluster 
catalog. 

Fig.~\ref{fig13} provides a statistical summary of the mask by showing the lower
limit to the X-ray luminosity for a detection as a function of redshift
for the median sensitivity of the sky and for the 10\% region with the lowest
sensitivity. Similar curves for other percentiles comprising the area of $\le 70\%$
of the best exposure in the sky are practically indistinguishable from the 
median line.
 
From the mask we can also calculate the effective survey volume of {\sf REFLEX II} 
as a function of X-ray luminosity, which is the basic statistic on
which for example the construction of the luminosity function depends. This
statistic is shown in Fig.~\ref{fig14} for the fiducial $\Lambda$CDM cosmological
model defined in the introduction. The calculations were limited to redshifts
up to $z = 0.8$ with the results given by the solid line, and further
calculations with redshift limits of $z = 0.3$ and $z = 0.22$ are also shown
by dashed lines in the Figure.
For a luminosity of $L_X = 5 \times 10^{44}$
erg s$^{-1}$, which is not far from $L^{\star}$ of the X-ray luminosity function,
we reach a survey volume of $\sim 2.5$ Gpc$^3$. The redshift limit of $z = 0.3$
approximately correspond to this limiting luminosity (as seen in Fig.~\ref{fig13}).

\section{ Statistical properties of the cluster sample}

The cluster number counts per unit sky area as a function of limiting flux,
the so-called logN-logS distribution, for the {\sf REFLEX II} cluster 
sample for 861 clusters detected with more than 20 counts is shown in 
Fig.~\ref{fig15}. The source density is calculated with a sky area normalization 
derived from the nominal flux, $F_n$, and the sensitivity map shown in 
Fig.~\ref{fig10}. The flux values are the corrected observed flux, $F_X$ 
(solid line), and the flux corrected for missing flux (dashed line). The best
slope for the observed LogN-logS function is about $-1.39$ (see
also Fig A.1 in the Appendix).

To account for selection biases in the presence of flux errors in
the determination of the logN-logS function we model the underlying function
as a power law. We convolve this function with the flux error and fit
the resulting function to the data by means of a maximum likelihood
method (e.g. Murdoch et al. 1973). 
We use two models for the flux error. In the first approach we
assume a constant error of 20\% and in the second model we use an error
that is decreasing with increasing flux according to $32.8\% \times F_X^{-0.505}$
which is motivated by Eq. (7) for a mean exposure time of about 360 sec.
In the constant error model we determine a function slope of $1.405 (\pm 0.075)$
and in the variable error model we find a shallower slope of $1.36  (\pm 0.07)$.
The uncertainty in the normalization is about 7 - 8\%. Among the two models
applied we consider the one with variable flux error as the more precise description.

The cluster distribution in luminosity redshift space is shown in Fig.~\ref{fig16}.
97 clusters have redshifts above $z = 0.25$.
In the figure we marked clusters listed in the Abell catalogs (Abell 1958,
Abell et al. 1989). In total 421 of the {\sf REFLEX II} clusters are from Abell's
sample. There are only 18 Abell clusters above redshift 0.25 and 4 above $z=0.3$.
The median redshift of the {\sf REFLEX II} is $z = 0.102$.

In Fig. \ref{fig17} we show the cluster surface density in the sky as a function 
of interstellar hydrogen column density, $n_H$. 
For this calculation we normalized the number of galaxy clusters in each bin
by the sky area with this $n_H$ range. For all sky regions, where the local
flux limit is smaller than the nominal flux limit, we apply a correction factor
determined from the logN-logS function as $R = {N(>F_{lim~nominal}) \over N(>F_{lim~local})}$.
We devide all sky regions where R is not unity by R and use this "effective"
sky area for the normalization. We note in Fig. \ref{fig17} that there 
is no variation of the detection efficiency as a function of interstellar
absorption. An exception is the last bin, in which all regions with $n_H$-values
above $10^{21}$ cm$^{-2}$ are collected. Only in this bin do we observe a 
small deficit.

\section{Comparison to other cluster surveys}

In this section we compare our cluster detections to some previous 
cluster surveys in order to check the completeness of our catalog.
We include here four major surveys, the published cluster detections of
the {\sf PLANCK} microwave satellite (PLANCK-Collaboration 2011), which detects 
clusters through the hot intracluster plasma via the Sunyaev-Zel'dovich effect, 
the MACS survey (Ebeling et al. 2000) based on the RASS, the 400d Survey
(Burenin et al. 2007), which is based on {\sf ROSAT} pointed observations,
and the cluster sample from the South Pole Telescope Sunyaev-Zel'dovich effect
survey published by Reichardt et al. (2012). 
We also check the consistency with the SGP cluster
sample of Cruddace et al. (2002).

While 83 of the clusters in the catalog of the early data release (PLANCK-Collaboration 2011) 
coincide with clusters in the {\sf REFLEX II} catalogue, nine {\sf PLANCK} clusters in
the {\sf REFLEX} II area have not been indentified in our cluster selection process.
PLCK277.8-51.7, is below the {\sf REFLEX II} flux limit, three (PLCK345.4-39.3, PLCK287.0-32.9,
PLCK 262.7-40.9) fall into very low exposure regions and have no significant 
detections in the RASS, one with intermediate exposure, PLCK292.5-22.0, 
has only 11 photons. The remaining four clusters that should be included
in the {\sf REFLEX II} catalog, PLCK206.0-39.5, PLCK239.3-26.0, PLCK308.3-20.2,
and PLCK283.2-22.9 have redshifts in the range 0.39 to 0.44 (with one redshift
unknown) and were missed by the limited depth of our optical identification. 
Four other southern clusters, PLCK288.6-37.7, PLCK271.2-31.0, PLCK286.6-31.3, 
and PLCK304.8-41.4 fall into the regions of the Magellanic clouds that 
have been cut out from our survey.

\begin{figure}
\includegraphics[width=\columnwidth]{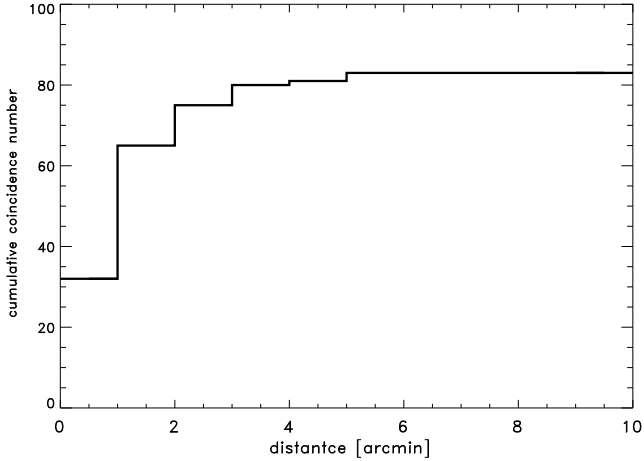}
\caption{Angular distance between the {\sf PLANCK} and {\sf REFLEX II} detections of 
83 clusters overlapping in the two catalogs. 78\% of the clusters
are found with a detection offset smaller than 2 arcmin. 
}\label{fig18}
\end{figure}

We also studied the positional coincidence of the {\sf PLANCK} and {\sf REFLEX II}
detections. The result is shown in Fig. \ref{fig18}. Of the total number
of 83 overlapping clusters 78\% are found within a matching radius of
2 arcmin. This is not too surprising, even though the positional uncertainty
for the {\sf PLANCK} detections is about 5 arcmin, as X-ray information was also
used in the validation and compilation of the clusters detected by {\sf PLANCK}.

Comparing our results to the published catalogs of the brightest (Ebeling et al. 2010)
and the most distant MACS clusters with $z > 0.5$ (Ebeling et al. 2007)
we find that only two clusters RXCJ0159.8-0849 ($z = 0.406$) and RXCJ0547.0-3904
($z = 0.319$) have not been included into our sample prior to the 
publication of the MACS data, because we had no clear sign of a cluster 
in optical images and we had not performed follow-up observations yet. 
But they had been marked as unidentified objects that could possibly be clusters
and are now included in the {\sf REFLEX II} sample.
All other 22 clusters in the {\sf REFLEX} area of the bright sample,
and the 2 clusters of the distant sample, RXCJ0454.1-0300 and RXCJ2214.9-1359
with fluxes above our survey limit have been detected in {\sf REFLEX II}. 

Comparing to the cluster catalog constructed from {\sf ROSAT} pointed observations by 
Burenin et al. (2007) we find that 101 clusters fall into the {\sf REFLEX} area.
Only one cluster that should most probably be in {\sf REFLEX II} has been missed since the
X-ray flux of this X-ray source is by far dominated by a star and the 
X-ray emission cannot be deblended with the angular resolution of the RASS.
This cluster, RXCJ1501.3-0830, has not been included in the present catalog.

We have also compared the galaxy cluster X-ray sources of our catalogue with 
the South Galactic Pole (SGP) X-ray cluster sample by Cruddace et al. (2002) 
which resulted from a precursor project of {\sf REFLEX}. 
We find that our catalogue recovers, as expected, all the sources in the 
South Galactic Pole sample above the flux limit. There are five clusters 
in the SGP catalog listed with a higher flux than the flux limit used by us:
RXCJ0012.9-0853, RXCJ0108.5-4020, RXCJ0213.9-0253, RXCJ0251.7-4109, 
RXCJ2346.7-1028. These sources have been found in our detailed analysis
of the X-ray emission to contain contamination from a second, in most cases
significantly softer X-ray source. These sources fall after deblending of the
second source below the flux limit of {\sf REFLEX II} and have been excluded from
our sample. Another cluster source, RXCJ2214.4-1701, has an AGN in the center of
the cluster and is consistent with a point source. We have therefore also removed
this object from the {\sf REFLEX II} catalog.

The final comparison is with the galaxy cluster catalogue from the 720 deg$^2$ 
area survey by the South Pole Telescope (SPT) in the millimeter regime detecting
the clusters through the Sunyaev-Zel'dovich effect (Reichardt et al. 2012).
Of the 224 galaxy clusters in the catalogue 13 are in the {\sf REFLEX II} sky
and flux limit. 12 are contained in the catalog and one, RXCJ2332.3-5358, 
is at the flux boundary and has a redshift of $z=0.4020$. As the flux is just 
2\% above the flux limit in our recent analysis, we include the cluster in our sample.
Again we find that no cluster with a redshift below $z = 0.35$ has been missed
by our compilation.

\section{Discussion}

\begin{figure}
\includegraphics[width=\columnwidth]{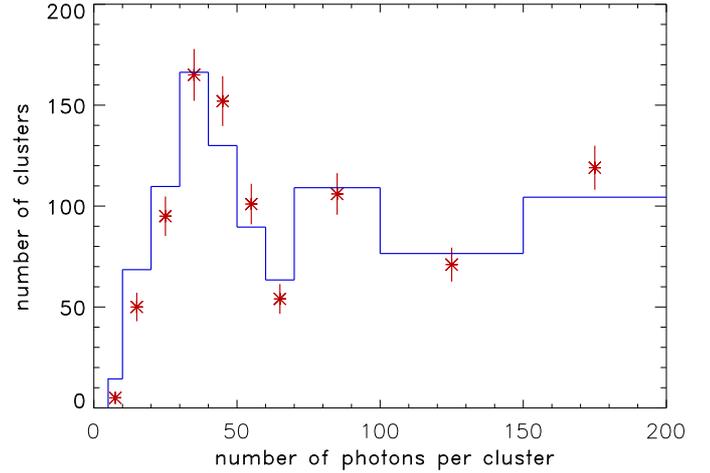}
\caption{Number of clusters expected in bins of photon number as calculated
from the best fit logN-logS power law function (blue solid line)
compared to the observed galaxy clusters sorted into bins of
detected number of photons.
}\label{fig19}
\end{figure}

Performing an X-ray survey of galaxy clusters
constitutes a special challenge as the source identification 
process is more complex than for most other surveys. First, the 
counterpart of an X-ray source in the optical is not a single object coinciding
with the source center, but a collection of galaxies which have to be proven to 
be close in redshift space. Second, it has to be shown that the X-ray emission of the
detected source originates from the intracluster medium of the cluster and not from 
another X-ray source in the line-of-sight of a cluster or an AGN in the cluster. In this 
case it crucially helps to detect the X-ray emission of the source as extended. This
is in our case true for 69\% of the {\sf REFLEX II} clusters, but we cannot rely
on this property for the remaining fraction of {\sf REFLEX II} catalog objects. Last,
for providing proper X-ray parameters and to justify the inclusion of a cluster
in the flux-limited catalog we have to make sure that the X-ray emission is not
substantially contaminated by another X-ray source (in particular from an AGN in
the cluster). This makes the X-ray source identification, as sketched in Fig.~\ref{fig7}
very complex. For the current catalog we can give a very high probability that most of
the cluster sources have a correct cluster identification. But improving the catalog
is to some degree a continuing effort. We therefore keep a detailed record on any
uncertainties of the identification process and use various sources of new information
(as for example deeper pointed observations) to improve the quality of the
{\sf REFLEX II} sample.

An important quality criterion of a survey is its completeness within the given selection
parameters. There is no obvious way to test the completeness of our catalog in an 
independent way. But we can perform a number of internal consistency checks and we
can compare to other cluster surveys that cover part of the survey parameter space
of {\sf REFLEX} to see which objects might have been missed.

A first obvious check is to look for any deficiency of galaxy clusters as 
a function of X-ray flux. The source identification becomes more difficult 
with decreasing flux, not only because there are in general fewer photons for 
the detection and the further characterization of the X-ray source, but 
also other means of identification like the inspection of optical images 
becomes notably more difficult on average at lower fluxes. Therefore we 
expect that these deficiencies would show up most significantly at the 
lowest fluxes. The good fit of a straight power law line to the logN-logS
distribution shown in Fig.~\ref{fig15} already indicates that there 
is no severe deficiency. 

To quantify this fact we use the fit of a power law to the number count function 
in Fig.~\ref{fig20} in the Appendix to test for deviations at the low luminosity end
and find a small deficit of $45 \pm 29$ clusters. Restricting the fit to a higher flux
range of $F_X \ge 3 \times 10^{-12}$ erg s$^{-1}$ cm$^{-2}$ results in a slightly steeper
slope of $-1.44$ yielding a formal deficit of $110 \pm 30$ clusters. 
This does not take into account, however, that the logN-logS 
distribution is not expected to be a straight power law but is expected to bend
towards a shallower slope. Thus this estimate is a pessimistic upper limit.
A more correct calculation can be performed when the X-ray luminosity function
and its evolution has been determined.

In a second test we inspect the photon number count distribution shown in Fig.~\ref{fig19}. 
To determine this distribution we use the best fit to the logN-logS function to
predict the expected number of clusters for each sky pixel with given detection sensitivity
with photon counts in ten bins of photon number. The bin boundaries for the ten bins
are 5, 10, 20, 30, 40, 50, 60, 70, 100, 150, infinity. In Fig.~\ref{fig19} we compare this 
predictions with the actually observed photon number distribution. The two distributions
agree quite well. Inspecting the low photon count region we find lower numbers for
the observations than for the prediction: 95 clusters in the bin 20 - 30 photons
for a predicted number of $110 \pm 10$ counts, 50 instead of 68.5 counts in the bin
10 - 20 photons, and 5 instead of 14.4 in the bin 5 to 10 photons. There is no
deficit in the next three higher bins. Therefore
the possible missing number of clusters in the lowest three bins is about 43.
For the cluster selection with a minimum number of 20 photons the $1\sigma$
upper limit of the deficit is 25 clusters. 

For the last test we use the $N_H$ distribution in Fig.~\ref{fig17}
which shows no significant density variation with interstellar absorption,
except for the last bin. The deficit in the last bin amounts to about 
21 clusters which is a small fraction of the total sample.

The comparison with other catalogs leads to the following implications. 
In the comparison with the {\sf PLANCK} detections where 4 out of 87 clusters 
were not found in the {\sf REFLEX II} catalog, we conclude that $6.4\%$ have been 
missed. These missing clusters all have high redshifts $>0.3$, a distance range
for which we do not claim a high completeness. Comparing to MACS we
find 2/24 cluster missing which yields a deficit of $8.3\%$. Again this 
concerns mostly the high redshift clusters.
The comparison with the 400 degree survey yields 1/101 
missing which is a fraction of $\sim 1\%$.

In summary none of the tests have shown a serious deficit larger than 10\%.
We expect a larger fractional deficit at high redshifts ($z \ge 0.35$), 
where it gets more difficult to identify all clusters without much more 
follow-up observations. Our best guess for the total number of missing 
clusters is thus of the order of about 50. 
 
For the test of the possible fraction of clusters severely contaminated by AGN X-ray
emission we refer the reader primarily to the inspection of the hardness ratio
distribution as shown for {\sf REFLEX I} (B\"ohringer et al. 2004) where we concluded that
the contamination should not be larger than about 6\%. A similar exercise with the  
current catalog gives contamination fractions not much larger.
Therefore we conclude that the clusters where contamination by AGN may have been
overlooked is not larger than 10\% and our best guess for the number of clusters
with contamination problems is also of the order of about 50. 

\section{Summary and Conclusion}

The {\sf REFLEX II} sample with 915 galaxy clusters together with the
northern {\sf NORAS II} sample of RASS detected clusters constitutes the largest 
statistically well defined sample of X-ray luminous galaxy clusters to date and
will probably remain so until the exploitation of the {\sf eROSITA} All-Sky X-ray Survey 
(Predehl et al. 2011) becomes effective. 
The sample will therefore be important for astrophysical as well
as cosmological studies. For these investigations the statistical properties of the
survey and the survey selection process has to be well known. We have therefore made
a large effort to construct a detailed three-dimensional survey selection function
providing the luminosity selection limit as a function of the sky position and 
redshift. This provides a basis not only for the construction of simple zero order
distribution functions like the X-ray luminosity function but also that of higher order
functions like the N-point correlation functions, the power spectrum or e.g. Minkowski
functionals. 

The survey selection function is not the only ingredient for the modeling of the
survey in cosmological applications. A precise characterization of the measurement 
errors is equally important. Here it is the uncertainty of the measured photon counts
which is practically proportional to the error of the measurement of the source 
flux and X-ray luminosity. As an 
improvement over {\sf REFLEX I} where we have used a mean flux error for the modeling, we
have made an effort to derive a more detailed description of the flux uncertainty.
Through a careful study of the dependence of the flux error on various cluster parameters,
we have found that the most important parameter dependence of the flux error can be
modelled as a function of $flux \times exposure$ as described in section 3. 

We have shown in section 8 that the best estimate for the completeness of the
{\sf REFLEX II} catalog is of the order of 95\% while we expect a contamination level of
about 5\%. Therefore it is not surprising that with this good quality of the catalog
we can obtain a good, uncontaminated measure of the power spectrum of the spatial
distribution of the clusters (Balaguera-Antolinez et al. 2010). 

To estimate how many clusters are still waiting to be detected in the
{\sf REFLEX} area of the RASS among the more than 100 000 X-ray sources in total, we
can use the same type of calculation as used for the photon number distribution
shown in Fig.~\ref{fig17}. We estimate the total number of clusters expected to have more
than six counts without any flux limit. With the low X-ray background of the RASS,
a detection of six photons has still a high probability to be real. The result for
the total number of these clusters in the {\sf REFLEX} region is about 9100. Thus we have only
detected a small fraction of all clusters in the RASS so far. We have, however,
reached some quality limit. Pursuing cluster detection to lower fluxes would increase
the mean flux error to values higher than 20\% and also the cluster identification
is getting notecibly more difficult.

\begin{acknowledgements}
We like to thank the {\sf ROSAT} team at MPE for the support with the data
and advice in the reduction of the RASS and the staff at ESO La Silla
for the technical support during the numerous observing runs conducted
for the {\sf REFLEX} project since 1992. Special thanks goes to Peter Schuecker,
who was a very essential part of our team and who died unexpectedly in 2006.
H.B. and G.C.  acknowledge support from the DfG Transregio Program TR33,
the Munich Excellence Cluster ''Structure and Evolution of the Universe'',
and under grant no. 50 R 1004 by the DLR.  

\end{acknowledgements}

\appendix
\section{LogN-LogS distribution for the nominal flux}

\begin{figure}
\includegraphics[width=\columnwidth]{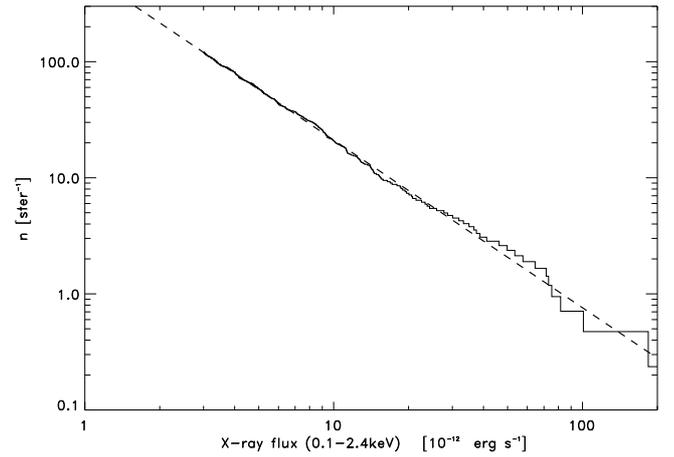}
\caption{Cumulative number counts of the {\sf REFLEX II} clusters as a
function of limiting flux per steradian for the nominal flux, $F_n$.
To characterize this function analytically we fitted a power-law function
to the data, as shown in the figure, obtaining a value for the slope of
 $1.385 (\pm 0.075)$.
}\label{fig20}
\end{figure}

The nominal flux defined and used for the construction of the 
{\sf REFLEX II} cluster sample is directly related to the observed
number counts and only depends on the interstellar hydrogen absorption
column density, $n_H$. It is therefore a purly observational 
quantity analogous to an optical magnitude corrected for extinction.
Therefore the logN-logS distribution for the nominal flux is an
important statistic for the characterization of the {\sf REFLEX II}
survey and it has been used in sections 5 and 8 for various calculations.
Therefore we show this function in Fig.~\ref{fig20}.

\end{document}